\documentclass{article}

\usepackage[utf8]{inputenc}
\usepackage{amsmath}
\usepackage{mathtools}
\usepackage{amsfonts}
\usepackage{physics}
\usepackage{float}
\usepackage{graphicx}
\usepackage{verbatim}

\usepackage{bbm}
\usepackage{bm}
\usepackage{algorithmic}
\usepackage{algorithm}
\usepackage[round]{natbib}
\usepackage{amsthm} 
\usepackage{algorithmic}
\usepackage{algorithm}
\usepackage{geometry}
\usepackage{mleftright }
\usepackage{tabularx}
    \newcolumntype{L}{>{\raggedright\arraybackslash}X}
\usepackage{epstopdf}
\usepackage[caption=false]{subfig}

\usepackage[round]{natbib}

\usepackage{amscd,amsfonts,amsmath,amssymb,amsthm,bbm,bm,latexsym,mathrsfs}
\usepackage{epsfig,graphics,graphicx,longtable}
\usepackage[english]{babel}
\usepackage[usenames]{color}
\usepackage{bookmark}
\usepackage{sgame}
\usepackage{gensymb}
\usepackage{epstopdf}
\usepackage{float}
\usepackage{comment}

\usepackage{tikz}
\usetikzlibrary{matrix}

\usepackage{geometry}

\makeatletter
\newcommand{\svast}{\bBigg@{3}}
\newcommand{\vast}{\bBigg@{4}}
\newcommand{\Vast}{\bBigg@{5}}
\makeatother

\begin{document}


\title{\bfseries \sffamily A Bayesian Non-linear State Space Copula Model to Predict Air Pollution in Beijing}
	\date{\small \today}
			\author{Alexander Kreuzer\footnote{Corresponding author: {E-mail: a.kreuzer@tum.de}} \footnotemark[3], Luciana Dalla Valle\footnotemark[2], and Claudia Czado\footnotemark[3]}
\date{%
	Technische Universit\"at M\"unchen\footnotemark[3] and University of Plymouth\footnotemark[2]\\[2ex]%
	\today
}
			\maketitle
\vspace*{-0.2cm}


\begin{abstract}

Air pollution is a serious issue that currently affects many industrial cities in the world and can cause severe illness to the population.
In particular, it has been proven that extreme high levels of airborne contaminants have dangerous short-term effects on human health, in terms of increased hospital admissions for cardiovascular and respiratory diseases and increased mortality risk.
For these reasons, accurate estimation and prediction of airborne pollutant concentration is crucial.
In this paper, we propose a flexible novel approach to model hourly measurements of fine particulate matter and meteorological data collected in Beijing in 2014.
We show that the standard state space model, based on Gaussian assumptions, does not correctly capture the time dynamics of the observations.
Therefore, we propose a non-linear non-Gaussian state space model where both the observation and the state equations are defined by copula specifications, and we perform Bayesian inference using the Hamiltonian Monte Carlo method.
The proposed copula state space approach is very flexible, since it allows us to separately model the marginals and to accommodate a wide variety of dependence structures in the data dynamics.
We show that the proposed approach allows us not only to predict particulate matter measurements, but also to investigate the effects of user specified climate scenarios.
\end{abstract}

\section{Introduction}\label{intro}

Over recent decades, rapid economic development and urbanization lead to severe and chronic air pollution in China, which is currently listed as one of the most polluted countries in the world.
Airborne pollutants contribute not only to the contamination of the air, but also of food and water, making inhalation and ingestion the major routes of pollutant exposure,  in addition to dermal contact, to a minor extent (\cite{kampa2008human}).
Exposure to ambient air pollution has been associated with a variety of adverse health effects, ranging from cardiovascular and respiratory illnesses, such as stroke and ischemic heart disease, to cancer and even death. Human health effects  include birth defects, serious developmental delays in children, and reduced activity of the immune system, leading to a number of diseases (\cite{liang2015assessing}). It has been shown that air pollution increases mortality and morbidity and shortens life expectancy (\cite{world2013review}), with heavy consequences in terms of health care and economy (\cite{song2017air}).
Outdoor PM2.5 has been established as one of the best metrics of air pollution-related risk to public health, since it is considered to be the fraction of air pollution that is most reliably associated with human disease (\cite{liu2017spatial}). In particular, PM2.5 is known to be a better predictor for acute and chronic health effects than other types of particulate matter pollutants (\cite{matus2012health}).
PM2.5 consists of fine particulate matter with aerodynamic diameters of less than 2.5 micrometers ($\mu$m).
PM2.5 is a portion of air pollution that is made up of extremely small particles and liquid droplets containing acids, organic chemicals, metals, and soil or dust particles, that are able to travel deeply into the respiratory tract, reaching the lungs. Sources of PM2.5 include combustion in mechanical and industrial processes, vehicle emissions, and tobacco smoke.
It has been estimated that in China, ambient PM2.5 was the first-ranking mortality risk factor in 2015 and exposure to this pollutant caused 1.1 million deaths in that year (\cite{cohen2017estimates}).

Fine particulate matter is a key driver of global health and therefore it is vital to accurately model and estimate the exposure to PM2.5, especially in areas of severe and persistent air pollution such as China and its biggest cities like Beijing. An accurate estimation and forecast of air pollution is crucial for a realistic appraisal of the risks that airborne contaminants pose and for the design and implementation of effective environmental and public health policies to control and limit those risks (\cite{shaddick2018data}).

Most of the contributions in the literature focus on modeling the observed concentrations of ambient air pollution.
For example, \cite{sahu2006spatio} modeled fine atmospheric particulate matter data collected in the US using a Bayesian hierarchical spatio-temporal approach.
\cite{sahu2005bayesian} used a spatio-temporal process based on a Bayesian kriged Kalman filtering to model atmospheric particulate matter in New York City.
\cite{calder2008dynamic} adopted a Bayesian dynamic process convolution approach to provide space-time interpolations of PM2.5 and PM10 concentrations readings taken across the state of Ohio.
State space representations, in addition to modeling the observed concentrations of air pollution, allow us to obtain an estimate of underlying non-measured factors, which are critical to assess pollution-related health risks.  

In this paper, we propose a novel flexible non-linear non-Gaussian state space model based on copulas, that includes a dynamic latent smoothing effect. As opposite to traditional approaches, this model allows us to identify time-points where the latent states have a considerable impact on the response. These points correspond to unusual high levels of air pollution, which cannot be accommodated for simply by the model including covariate effects such as weather conditions and seasonal patterns.
These can have dangerous effects on human health.
Extreme air pollution levels need to be carefully monitored, since it is proven that acute exposures increase the rate of cardiovascular, respiratory and mortality events (\cite{anderson2012clearing}).
The negative effects of pollution spikes to human health are known since the twentieth century.
For example, in 1930 local factory emissions caused the formation of a dense fog over the Meuse Valley in Belgium. Over 3 days, several thousand people were stricken with acute pulmonary symptoms, and 60 people died of respiratory causes (\cite{nemery2001meuse}). 
Another well-known event is the great smog of London of 1952, when a dense haze descended upon the city, resulting in more than 3,000 excess deaths over 3 weeks and 12,000 through February 1953 (\cite{bell2001reassessment}).
Recent studies in various countries confirm the severity of short- and long-term effects of the exposure to increased levels of airborne contaminants on human health, including respiratory diseases, decreased lung functions, recurrent health care utilization, reduced life expectancy and increased mortality.
Vulnerable people, such as infants and elderly, are particularly susceptible to extreme air pollution levels. 
In particular, children who are exposed to an excess level of PM2.5 are under a significantly high risk of hospitalization for respiratory symptoms, asthma medication use, and reduced lung function, while PM2.5 pollution is linked to an increased risk of hospital admission for heart failure among the elderly.
In addition, air pollution has a substantial economic impact, since it multiplies the world wide healthcare burden (\citep{anderson2012clearing, kan2012ambient, kim2015review}).
We will show that our methodology performs better than traditional approaches in accurately modelling and predicting unusual high levels of air pollutants, allowing us to better assess the effects of human exposure to airborne contaminants.
Our statistical approach in contrast to just smoothing observations also allows to investigate the effect of changed climate conditions on the predicted PM2.5 levels. Illustrations of such climate simulations are also given.

\subsection{Linear Gaussian state space models}

State space models are dynamic statistical analysis techniques, which assume that the state of a system at time $t$ can only be observed indirectly through observed time series data (\cite{durbin2000time}). 
State space models contain two classes of variables, the unobserved state variables, which describe the development over time of the underlying system, and the observed variables (\cite{durbin2002simple}).  
The univariate linear Gaussian state space model with continuous states and discrete time points $t=1,\ldots,T$ can be formulated as follows
\begin{eqnarray}
Z_t & = & \rho_t^{obs} W_t + \sigma_t^{obs} \eta_t^{obs} \label{obs_eq_v1} \\ 
W_t & = & \rho_t^{lat} W_{t-1} + \sigma_t^{lat} \eta_t^{lat}.  \label{state_eq_v1}
\end{eqnarray}
Here, $(Z_t)_{t=1, \ldots, T}$ is a random vector corresponding to the observations, $(W_t)_{t=1, \ldots, T}$ is an unobserved state vector and $\eta_t^{obs}$ and $\eta_t^{lat}$ are independent disturbances, with $\eta_t^{obs} \sim \mbox{N}(0,1)$ and $\eta_t^{lat} \sim \mbox{N}(0, 1)$ for $t=1, \ldots, T$. Further, it holds that $\rho_t^{obs} \in (-1,1)$ , $\rho_t^{lat} \in (-1,1)$, $\sigma_t^{obs} \in (0, \infty)$ and $\sigma_t^{lat} \in (0,\infty)$. It is also assumed that $W_0 \sim \mbox{N}(\mu_{0}^{lat}, (\sigma_0^{lat})^2)$ is independent of $\rho_t^{lat}$ and $\rho_t^{lat}$ for all $t$, where $\mu_{0}^{lat}$ and $\sigma_0^{lat}$ are generally known.
Equation (\ref{obs_eq_v1}) is commonly referred to as the observation equation and it describes how the observed series depends on the unobserved state variables $W_t$ and on the disturbances $\eta_t^{obs}$. 
Equation (\ref{state_eq_v1}) is referred to as the state equation and it describes how these state variables evolve over time (\cite{van2010intervention}).

The linear Gaussian state space model can also be expressed as
\begin{eqnarray}
Z_t \, | \, W_t & \sim & \mbox{N} \left( \rho^{obs}_t W_t \, ; \,\, (\sigma_t^{obs})^2 \right) \label{obs_eq_v2} \\ 
W_t \, | \, W_{t-1} & \sim & \mbox{N} \left( \rho^{lat}_t W_{t-1} \, ; \,\, (\sigma^{lat}_t)^2 \right) \label{state_eq_v2}
\end{eqnarray}
Typically, Kalman filter recursions are used for determining the optimal estimates of the state vector $W_t$ given information available at time $t$ (\cite{durbin2012time}).
Other methods, such as Empirical Bayes was proposed by \cite{koopman2017empirical} to efficiently estimate dynamic factor models defined by latent stochastic processes, adopting a shrinkage-based approach.
\cite{ippoliti2012space} used a linear Gaussian state space model to produce predictions of airborne pollutants in Italy and in Mexico.

\subsection{Beijing ambient air pollution data}
\label{seq:data}
\begin{figure}[H]
\centerline{%
\includegraphics[trim={0 3cm 0 0},width=1\textwidth]{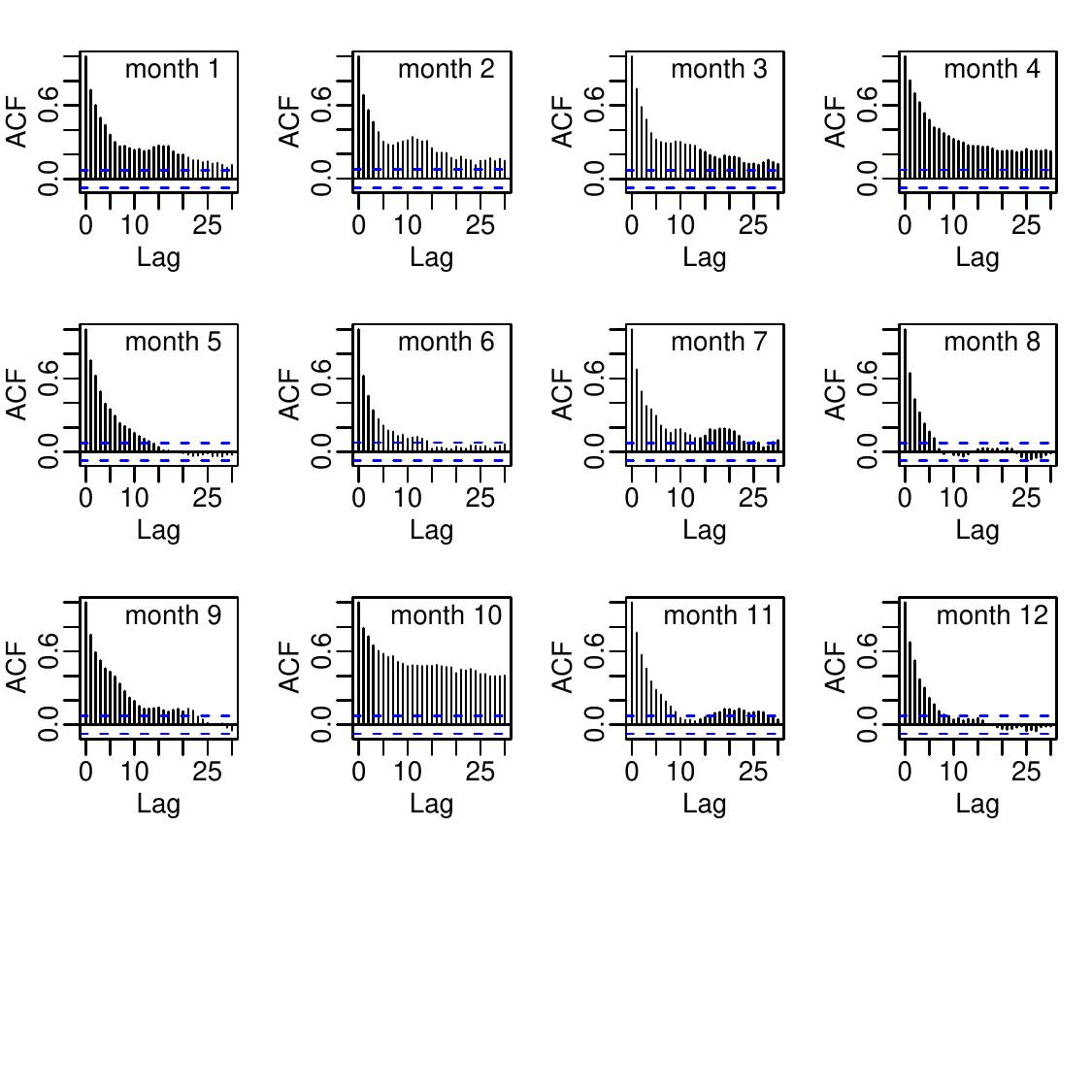}%
}%
\caption{Autocorrelation functions (acf) of $(\hat z_t)_{t=1, \ldots, T}$ for all 12 monthly data sets.}
\label{fig:acf}
\end{figure}

In this paper, we aim at accurately estimating and predicting the concentration of airborne particulate matter using a flexible state space model. 
We consider a dataset of hourly PM2.5 readings ($\mu$g$/\mbox{m}^3$) and meteorological
 measurements, such as dew point (DEWP, degrees Celsius), temperature (TEMP, degees Celsius), pressure (PRES, hPa), wind direction (CBWD, taking values: northwest (NW), northeast (NE), southeast (SE) and calm and variable (CV)), cumulated wind speed (IWS, m/s) and precipitations (PREC), collected in Beijing in 2014, and we split the data into 12 monthly sub-sets  \footnote{The dataset used in this paper is part of a larger dataset collected in Beijing during a 5-year time period, from January 1st, 2010 to December 31st, 2014, for a total of 43,824 observations. The data are available at \url{https://archive.ics.uci.edu/ml/datasets/Beijing+PM2.5+Data} }
 (\cite{liang2015assessing}). 
 This allows us to adjust the model over time periods.
In order to consider the effects of meteorological conditions on airborne contaminants concentrations, we assume a generalized additive model (GAM) (\cite{hastie1986generalized}).
More precisely, we suppose that, for each month, the relationship between PM2.5 concentrations $Y_t$ and covariates $\boldsymbol{x}_t$ for each hourly data point $t = 1, \ldots, T$ (where $T$ is the total number of monthly observations) is described by a GAM, such that 
\begin{equation} \label{eq:GAM}
Y_t = f(\boldsymbol{x}_t) + \sigma \varepsilon_t,
\end{equation} 
where $\boldsymbol{x}_t$ contains the meteorological covariates and seasonal covariates capturing within-day and -week patterns. Further $f(\cdot)$ is a smooth function of the covariates, expressing the mean of the GAM, and $\varepsilon_t \overset{\text{iid}}{\sim} N(0,1).$ For estimation we make use of the two step approach which is commonly used for copula models: we first estimate the GAM, fix the GAM parameters at point estimates, and then estimate the copula model (\cite{joe1996estimation}).
We define the standardized errors $Z_t$ as
\begin{equation} \label{eq:residuals}
    Z_t =  \frac{Y_t - f(\boldsymbol{x}_t)}{\sigma}, 
\end{equation}
for $t=1, \ldots, T$. 
This step allows us to account for weather and seasonal patterns. High values of $Z_t$ are then of interest to detect unusual high levels of pollution so far not accounted for.
Using the estimates $\hat{f}(\boldsymbol{x}_t)$ and $\hat{\sigma}$ of the GAM, we obtain approximately standard normal data $\hat{z}_t$ as
\begin{equation} \label{eq:hatresiduals}
    \hat{z}_t =  \frac{y_t - \hat{f}(\boldsymbol{x}_t)}{\hat{\sigma}},
\end{equation}
for $t=1, \ldots, T$.
The empirical autocorrelation function of $(\hat{z_t})_{t=1, \ldots, T}$ is shown for each month in Figure \ref{fig:acf}. We observe dependence among succeeding observations and therefore the independence assumption for the errors $\varepsilon_t$ of the standard GAM model in \eqref{eq:GAM} does not seem to be appropriate. We employ a state space model, as specified in \eqref{obs_eq_v1} and \eqref{state_eq_v1}, to allow for time effects in the GAM.
Here $\rho_t^{obs}$ and $\rho_t^{lat}$ will be estimated from the data. Further, we assume that they do not depend on time, i.e. we set $\rho_t^{obs}=\rho_{obs}$ and $\rho_t^{lat}=\rho_{lat}$.  
In our data application we split the data into monthly periods to make this assumption more plausible.

We now consider a state space model for $Z_t$, which is standardized by a GAM.
Under our assumptions we have $\sigma_t^{obs} = \sqrt{Var(Z_t | W_t)} = \sqrt{1 - \rho_{obs}^2}$ for $\rho_{obs} \in (-1,1)$ and $\sigma_t^{lat} = \sqrt{Var(W_t | W_{t-1})} = \sqrt{1 - \rho_{lat}^2}$ for $\rho_{lat} \in (-1,1)$. For the initial conditions we assume $\mu_0^{lat}=0$ and $\sigma_0^{lat}=1$.
 
With these assumptions the state space model in  \eqref{obs_eq_v1} and \eqref{state_eq_v1} becomes
\begin{equation}
\begin{split}
Z_t & = \rho_{obs} W_t + \sqrt{1 - \rho_{obs}^2} \eta_t^{obs}  \\ 
W_t & = \rho_{lat} W_{t-1} + \sqrt{1 - \rho_{lat}^2} \eta_t^{lat}  \label{state_eq_v1_res}
\end{split}
\end{equation}
with  $\eta_t^{obs}, \eta_t^{lat} \overset{\text{iid}}{\sim} N(0,1)$ and $W_0 \sim N(0,1)$.
Note that representation \eqref{state_eq_v1_res} induces the following bivariate normal distributions
\begin{equation*}
\begin{split}
\begin{pmatrix}
   Z_t \\
    W_t \\
\end{pmatrix} &\sim N_2\left( \begin{pmatrix}
   0 \\
    0 \\
\end{pmatrix} ,   \begin{pmatrix}
    1       & \rho_{obs}  \\
    \rho_{obs}       & 1 \\
\end{pmatrix}\right) \\
\begin{pmatrix}
   W_t \\
    W_{t-1} \\
\end{pmatrix} &\sim N_2\left( \begin{pmatrix}
   0 \\
    0 \\
\end{pmatrix} ,   \begin{pmatrix}
    1       & \rho_{lat}  \\
    \rho_{lat}       & 1 \\
\end{pmatrix}\right). \\
\end{split}
\end{equation*}

\begin{figure}[H]
\centerline{%
\includegraphics[trim={0 3cm 0 0},width=1.0\textwidth]{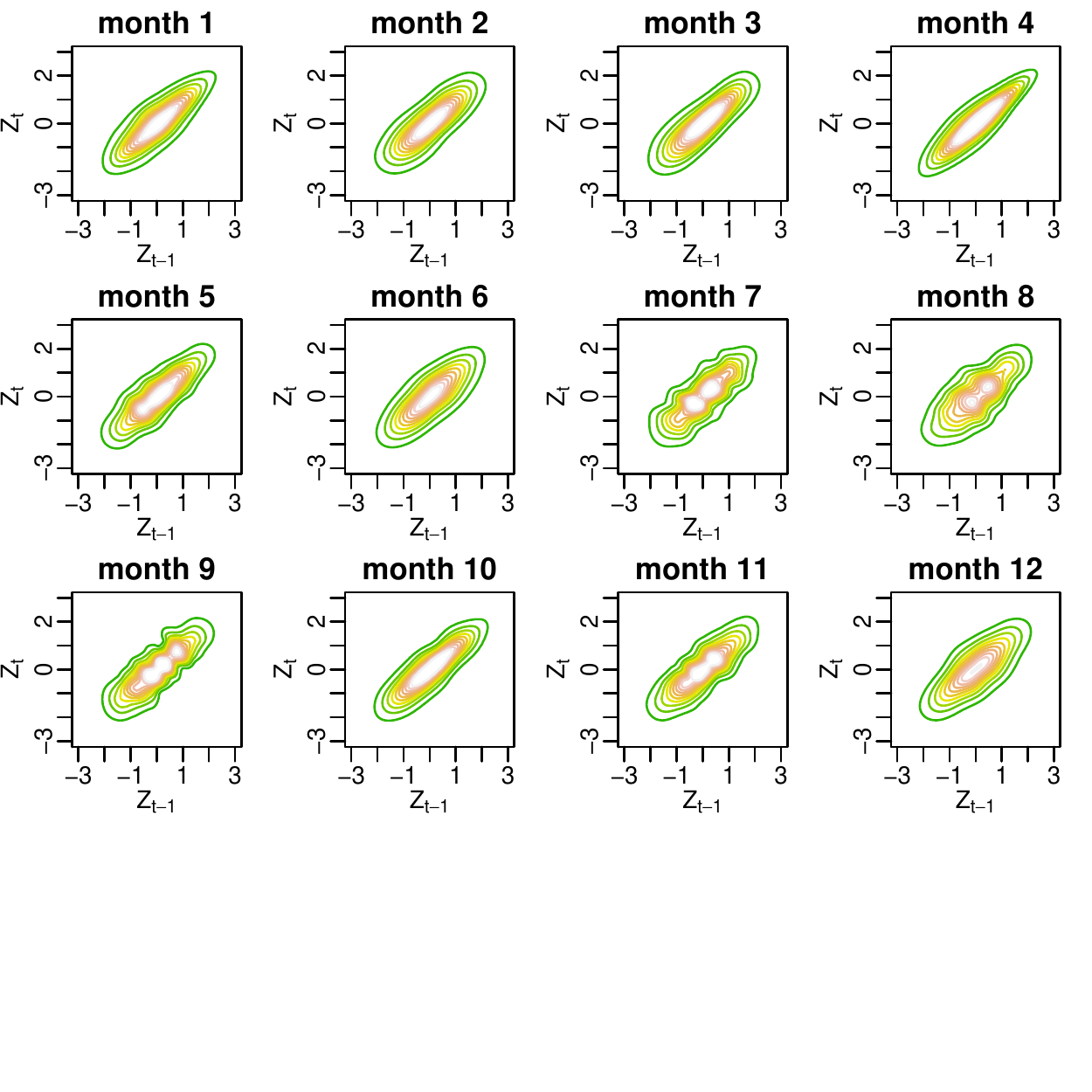}%
}%
\caption{Normalized contour plots of pairs $(\hat{z}_t, \hat{z}_{t-1})_{t=2, \ldots, T}$ ignoring serial dependence for each of the 12 Beijing air pollution monthly data sets.}
\label{fig:contour}
\end{figure}

In order to assess the suitability of the linear Gaussian state space model to the Beijing air pollution data, we display in Figure~\ref{fig:contour} the bivariate normalized contour plots of the pairs $(\hat{z}_t, \hat{z}_{t-1})_{t=2,\ldots,T}$ for each month, to visualize the dependence structure between two successive time points in the series. 
Using \eqref{state_eq_v1_res} we see that $Z_t$ can be written as a linear function of $Z_{t-1}$ and independent normally distributed disturbances. Since $Z_1$ is normally distributed, it follows that $(Z_t, Z_{t-1})$ are jointly normal.
In particular, we have 
$$
Z_t \sim N(0,1) \hspace{0.7cm}  \mbox{and} \hspace{0.7cm}  Cov(Z_t, Z_{t-1}) = \rho_{obs}^2 \rho_{lat} \hspace{0.7cm} \forall t \ge 1.
$$

However, Figure~\ref{fig:contour} reveals that the normalized contour plots of the Beijing monthly data deviate from the elliptical shape of a Gaussian dependence structure (which, to aid comparisons, is depicted in the top left panel of Figure~\ref{fig:cplots} in Appendix A). For example, the normalized contour plots for January and October (months 1 and 10) show tail dependence and/or asymmetry in the tails, which cannot be modeled with a Gaussian distribution.
This suggests that the linear Gaussian state space model is too restrictive for the Beijing air pollution data and a more flexible approach needs to be adopted.

\subsection{Our proposal}

In the literature, extensions of the linear Gaussian state space model, relaxing the assumptions of linearity and normality, have been studied, for example, by \cite{johns2005non}, who adopted a non-linear and non-Gaussian state space formulation to model airborne particulate matter, yet relying on the Normal distribution to describe the errors in the state and observation equations.
\cite{chen2012tracking} implemented a non-linear state space model to predict the global burden of infectious diseases using the extended Kalman filter approach. Non-linear state and observation equations of this model were derived from differential equations, however the authors still used Gaussian noise terms in the observation and state equations.

We propose a very flexible Bayesian non-linear and non-Gaussian state space model, where both the observation and the state equations are described by copulas.
First, we find an equivalent formulation of the Gaussian state space model in \eqref{state_eq_v1_res} in terms of copulas. The representation is given by
\begin{equation}
\begin{split}
(U_t \, , \, V_t)  & \sim  \mathbbm{C}_{U,V}^{Gauss} (\, \cdot \,, \, \cdot; \, \tau_{obs})  \\
(V_t \, , \, V_{t-1})  & \sim  \mathbbm{C}_{V_{2},V_{1}}^{Gauss} (\, \cdot \, , \, \cdot; \, \tau_{lat}), 
\label{eq:gaus_cop_ssm}
\end{split}
\end{equation}
where
\begin{equation} \label{eq:residuals}
    U_t = \Phi \left( Z_t \right), V_t = \Phi \left( W_t \right),
\end{equation}
with $\Phi$ denoting the standard normal cumulative distribution function. The variables $U_t$ and $V_t$ are marginally uniformly distributed on $(0,1)$ and $Z_t$ and $W_t$ are standard normal. Here  the Gaussian copulas $\mathbbm{C}_{U,V}^{Gauss}$ and $\mathbbm{C}_{V_{2},V_{1}}^{Gauss}$ are parametrized by Kendall's~$\tau$, obtained as $\tau_{obs}  =  \frac{2}{\pi} \arcsin (\rho_{obs})$ and 
$\tau_{lat} =  \frac{2}{\pi} \arcsin (\rho_{lat})$. Corresponding approximately uniform pseudo-copula data, that can be used for estimating the model in \eqref{eq:gaus_cop_ssm}, are obtained as 
\begin{equation} 
\label{eq:uresidual}
    \hat{u}_t = \Phi \left( \hat z_t \right). 
\end{equation}

By reformulating the state space representation in \eqref{state_eq_v1_res} in terms of copulas in \eqref{eq:gaus_cop_ssm}, it is straightforward to see how we can generalize the Gaussian linear state space model by replacing the Gaussian copulas in \eqref{eq:gaus_cop_ssm} with arbitrary bivariate copulas.
Typical restrictions of the Gaussian copula, such as symmetric tails, can be circumvented. For example, a Gumbel copula would allow for asymmetric tails.
\cite{koopman2016predicting} incorporated the symmetric-tailed Gaussian and Student t copulas in non-linear non-Gaussian state space models; however, asymmetric tail dependence could not be captured, since the authors ignored non-symmetric copula families and restricted their attention solely to autoregressive state equations. 
The proposed Bayesian copula-based state space model allows us to specify various types of dependence structures to model the relationships between the observations and the underlying states, and to describe the states evolution over time.   
We will show that our methodology is able to accurately model and predict the levels of PM2.5 in Beijing.

The remainder of the paper is organized as follows.
Section \ref{CopulaStateSpace} introduces a copula-based state space model, Section \ref{BayesAnalysis} illustrates the Bayesian inference for the proposed approach and Section \ref{DataAnalysis} is devoted to the application of the copula state space model to the Beijing pollution data. It also includes some simulations to study the PM2.5 predictions under different climate scenarios. Concluding remarks are given in Section \ref{Conclusions}.

\section{The copula state space model}\label{CopulaStateSpace}
\label{sec:model}
The copula state space model extends the linear Gaussian state space approach, allowing copula specifications in place of normal distributions as in the observation equation (\ref{obs_eq_v2}) as well as in the state equation (\ref{state_eq_v2}).
In particular, we assume that the dynamic behaviour of the residuals $Z_t := \Phi^{-1}(U_t)$ for the GAM model introduced in equation (\ref{eq:GAM}), with $Z_t \sim N(0,1)$ and $U_t \sim \mbox{U}(0,1)$ defined as in (\ref{eq:residuals}), depends on the latent variable $W_t := \Phi^{-1}(V_t)$, with $W_t \sim N(0,1)$ and $V_t \sim \mbox{U}(0,1)$, according to a bivariate copula distribution given in the observation equation.
The evolution of the latent variable $W_t$ over time is also described by a bivariate copula distribution, which defines the state equation of the model.
The copula distributions defining the observation and state equations of the proposed state space approach do not necessarily belong to the same family, allowing maximum flexibility in the specification of the model.
However, we restrict our model to bivariate copula families with a single parameter. This gives still a flexible class of copula families, including e.g. Gaussian, Gumbel, Clayton or Frank copulas. The Student t copula can also be included if we fix the degrees of freedom parameter. An overview of different bivariate copula families can be found in \cite{joe2014dependence}, Chapter 4. Further, we are able to express the copula dependence parameters in the observation and state equations in terms of Kendall's $\tau$. This is convenient for comparison of the dependence strength, since the parameter space of distinct copula families
may be different. 
More formally, we assume the following joint distributions for the uniformly transformed variables $U_t$ and $V_t$, with $t = 1, \ldots, T$ 
\begin{eqnarray*}
(U_t \, , \, V_t)  & \sim & \mathbbm{C}_{U,V}^{obs} (\, \cdot \,, \, \cdot; \, \tau_{obs})  \\
(V_t \, , \, V_{t-1})  & \sim & \mathbbm{C}_{V_{2},V_{1}}^{lat} (\, \cdot \, , \, \cdot; \, \tau_{lat}),  
\end{eqnarray*}
where $\tau_{obs} = g(\theta_{obs})$ is the Kendall's $\tau$ of the copula of the observations and $\tau_{lat} = g(\theta_{lat})$ is the Kendall's $\tau$ of the copula of the states (latent variables), respectively. The function $g$ is an appropriate one-to-one transformation function, and $\theta_{obs}$ and $\theta_{lat}$ are the parameters of the bivariate copulas $\mathbbm{C}_{U,V}^{obs}$ and $\mathbbm{C}_{V_{2},V_{1}}^{lat}$, respectively.  For the specification of $g$ for some one-parameter copula families see \cite{joe2014dependence}, Chapter 4.

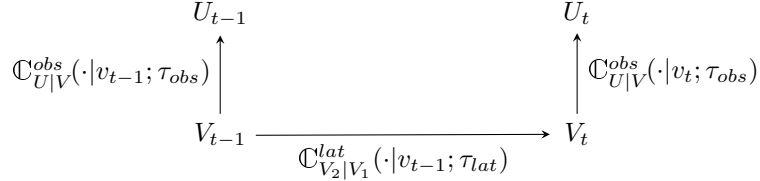
\begin{figure}[H]
\centering
\begin{tikzpicture}
  \matrix (m) [matrix of math nodes,row sep=3em,column sep=11em,minimum width=2em]
  {
     U_{t-1} &  U_{t} \\
     V_{t-1} &  V_{t} \\};
  \path[-stealth]
    (m-2-1) edge node [left] {$\mathbbm{C}_{U_{}|V_{}}^{obs} ( \cdot |  v_{t-1}; \tau_{obs})$} (m-1-1)
    (m-2-2) edge node [right] {$\mathbbm{C}_{U_{}|V_{}}^{obs} ( \cdot |  v_t;  \tau_{obs})$} (m-1-2)
    (m-2-1) edge node [below] {$\mathbbm{C}_{V_{2}|V_{1}}^{lat} ( \cdot  |  v_{t-1};  \tau_{lat})$} (m-2-2);
\end{tikzpicture}
\caption{Graphical visualization of the copula state space model.}\label{fig:cop_st_sp}
\end{figure}
\noindent

The copula state space model is defined on the uniform scale as follows
\begin{eqnarray}
U_t \, | \, V_t = v_t  & \sim & \mathbbm{C}_{U|V}^{obs} (\, \cdot \,| \, v_t; \, \tau_{obs})  \label{eq:obs_cop_2} \\
V_t \, | \, V_{t-1} = v_{t-1}  & \sim & \mathbbm{C}_{V_{2}|V_{1}}^{lat} (\, \cdot \, | \, v_{t-1}; \, \tau_{lat})  \label{eq:state_cop_2}
\end{eqnarray}
where  (\ref{eq:obs_cop_2}) is the observation equation and  (\ref{eq:state_cop_2}) is the state equation. We assume, as in the linear Gaussian state space model, that $U_t$ is independent of $U_{t-1}$ given the latent state $V_t$.
The copula state space model introduced in equations (\ref{eq:obs_cop_2}) and (\ref{eq:state_cop_2}) can be visualized as in Figure \ref{fig:cop_st_sp}.

\begin{figure}[H]
\centerline{%
\includegraphics[width=1\textwidth]{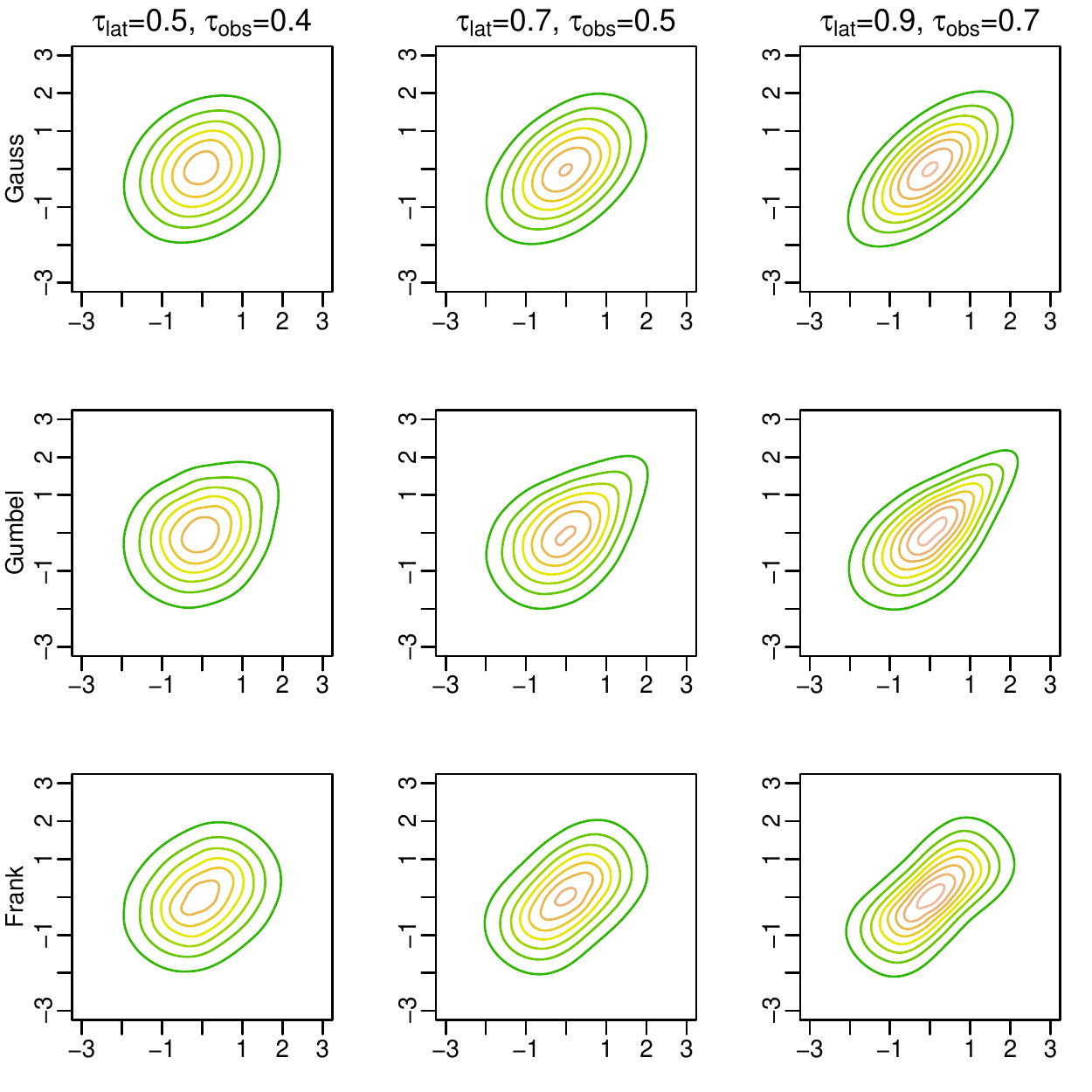}%
}%
\caption{Normalized contour plots for $(Z_t,Z_{t-1})$ of the copula state space model for different choices of bivariate copula families. In the state and observation equation we choose the same copula family.}
\label{fig:model_sims}
\end{figure}

We now derive the joint distributions for the normalized variables $Z_t$ and $W_t$
\begin{eqnarray}
(Z_t \, , \, W_t)  & \sim & F_{Z_t, W_t} \label{eq:joint_obs_norm} \\
(W_t \, , \, W_{t-1})  & \sim & F_{W_{t},W_{t-1}}.  
\end{eqnarray}
By Sklar's theorem (\cite{sklar1959fonctions}), the distribution (\ref{eq:joint_obs_norm}) can be expressed as
\begin{eqnarray*}
F_{Z_t, W_t} (z_t, w_t)  & = & \mathbbm{C}_{U_{},V_{}}^{obs} (\, \Phi(z_t) \,, \, \Phi(w_t); \, \tau_{obs}) \\
  & = & \mathbbm{C}_{U_{},V_{}}^{obs} (\, u_t \,, \, v_t; \, \tau_{obs}).  
\end{eqnarray*}
Hence, 
\begin{eqnarray*}
F_{Z_t| W_t = w_t} (z_t | w_t)  & = & \frac{\partial}{\partial v_t} \mathbbm{C}_{U_{},V_{}}^{obs} (\, \Phi(z_t) \,, \, v_t; \, \tau_{obs}) \bigg|_{v_t = \Phi(w_t)} \\
 & = & \mathbbm{C}_{U_{} | V_{}}^{obs} (\, u_t \,| \, v_t; \, \tau_{obs})\bigg|_{u_t = \Phi(z_t), v_t = \Phi(w_t)} \\
 & = & \mathbbm{C}_{U_{} | V_{}}^{obs} (\, \Phi(z_t) \,| \, \Phi(w_t); \, \tau_{obs}).  
\end{eqnarray*}
Similarly,  
$$
F_{W_t| W_{t-1} = w_{t-1}} (\cdot | w_{t-1}) = \mathbbm{C}_{V_{2} | V_{1}}^{lat} (\, \Phi(\cdot) \,| \, \Phi(w_{t-1}); \, \tau_{lat}).  
$$
Therefore, the model can also be expressed on the normalized scale as follows
\begin{eqnarray}
Z_t \, | \, W_t = w_t  & \sim & \mathbbm{C}_{U_{}|V_{}}^{obs} (\, \Phi(z_t) \,| \, \Phi(w_t); \, \tau_{obs})  \label{eq:obs_cop_1} \\
W_t \, | \, W_{t-1} = w_{t-1}  & \sim & \mathbbm{C}_{V_{2}|V_{1}}^{lat} (\, \Phi(w_t) \, | \, \Phi(w_{t-1}); \, \tau_{lat}),  \label{eq:state_cop_1}
\end{eqnarray}
where  (\ref{eq:obs_cop_1}) is the observation equation and  (\ref{eq:state_cop_1}) is the state equation. Contour plots of $(Z_t,Z_{t-1})$ of this model for different choices of bivariate copulas are shown in Figure~\ref{fig:model_sims}, illustrating different shapes that the model can deal with.

The copula state space model has the advantage of allowing flexibility in the specification of the observation and state equations, and thus is able to accommodate a wide variety of dependence structures in the air pollution data dynamics.

In the standard GAM the errors are assumed to be independent. 
Our methodology allows us to account for autoregressive effects in the error through the underlying latent variable $\sigma W_t$, as defined on the original scale of the GAM residuals, or via the proxy $V_t$, on the uniform scale. These latent variables can be interpreted as non-measured autoregressive effects.

As we will see in Section \ref{InSample}, our model's flexibility allows us to detect extreme air pollution levels, where the response is more susceptible to the effect of the underlying latent variable. 
Capturing unusual air contaminant levels is very important, since  human exposure to pollution spikes have a substantial impact on general health, causing severe cardiovascular and respiratory illness, and increasing mortality.

\subsection{Identifiability constraints}

We notice some identifiability issues related to the model.
In particular, if $\tau_{obs} = 1$, the observed and latent variables are equivalent and hence the state equation becomes unnecessary. 
In addition, if $\tau_{lat} = 0$, then the latent variables $(V_{t})_{t=1, \ldots, T}$ at different time points become independent.
Therefore, we need to set identifiability constraints for the copula state space model by establishing a relationship between $\tau_{obs}$ and $\tau_{lat}$.
In order to do that, we notice that the dependence between two successive time points $U_{t-1}$ and $U_t$ is determined by both $\tau_{lat}$ and $\tau_{obs}$. 
The form of the correlation between $Z_{t-1}=\Phi^{-1}(U_{t-1})$ and $Z_t=\Phi^{-1}(U_{t})$ can be derived exactly when $\mathbbm{C}_{U_{},V_{}}^{obs}$ and $\mathbbm{C}_{V_{2},V_{1}}^{lat}$ are both Gaussian copulas.
Since in the Gaussian case the parameter of the observation equation copula is the correlation coefficient $\rho_{obs}$ and the parameter of the state equation copula is the correlation coefficient $\rho_{lat}$, then the correlation between $Z_{t-1}$ and $Z_t$ is $\mbox{cor} (Z_{t-1}, Z_t) = \rho_{obs}^2 \rho_{lat}$.
The higher the value of $\rho_{lat}$ the smoother the latent states are. 
Higher smoothness of the latent states induces a lower prediction uncertainty for the latent states.
To guarantee a certain degree of smoothness, we need to set $\rho_{lat}$ greater than some specific value and therefore impose $\rho_{obs} \leq \rho_{lat}$ in our approach. 

In particular, we assume the identifiability constraint in the Gaussian case
$$
\rho_{obs} = \rho_{lat}^c \hspace{0.5cm} \mbox{for some suitable value} \hspace{0.5cm} c \geq 1.
$$
In this case, the correlation between $Z_{t-1}$ and $Z_t$ becomes $\mbox{cor} (Z_{t-1}, Z_t) = \rho_{lat}^{2c + 1}$.
Transforming the correlation coefficients into Kendall's $\tau$, in the Gaussian case, we obtain the following relationships
$$
\tau_{obs} = \frac{2}{\pi} \arcsin (\rho_{lat}^c) \qquad \mbox{and} \qquad
\tau_{lat} = \frac{2}{\pi} \arcsin (\rho_{lat}),
$$
hence, $\tau_{obs}$ is a function of $\tau_{lat}$ and $c$.

\begin{figure}[H]
\centerline{%
\includegraphics[width=0.6\textwidth]{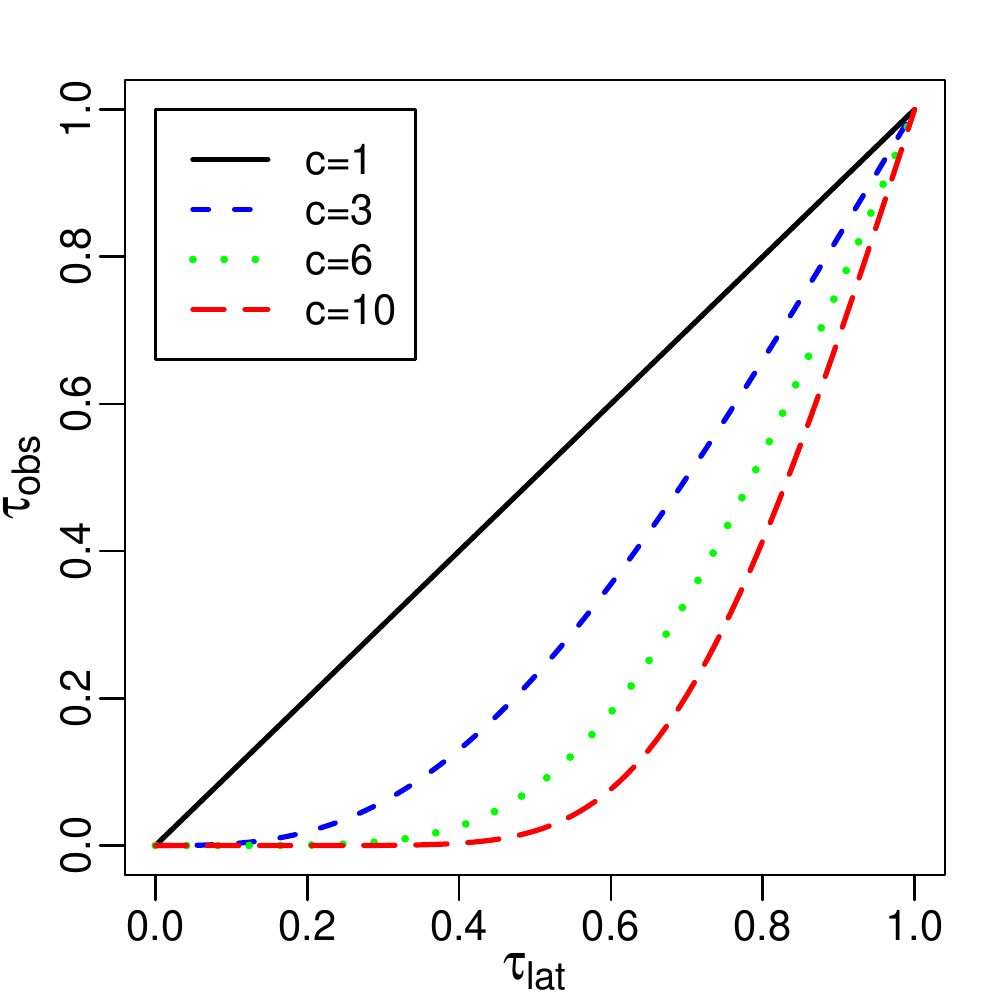}%
}%
\caption{Graphical representation of the relationship between the parameter $\tau_{obs}$ ($y$-axis) plotted against $\tau_{lat}$ ($x$-axis) in the Gaussian case for different values of $c = 1,3,6,10$.}
\label{fig:taurel}
\end{figure}

Figure \ref{fig:taurel} visualizes the relationship between the parameter $\tau_{obs}$ (on the $y$-axis) plotted against $\tau_{lat}$ (on the $x$-axis) in the Gaussian case for different values of $c = 1,3,6,10$.
Considering that the strength of dependence between $U_{t-1}$ and $U_t$ is increasing in $\tau_{lat}$ and in $\tau_{obs}$, Figure \ref{fig:taurel} shows that the higher the value of $c$ the higher $\tau_{lat}$ needs to be to achieve a fixed strength of dependence between $U_{t-1}$ and $U_t$. Therefore, for higher values of $c$ we expect to obtain a smoother behaviour of the latent states $(V_{t})_{t=1, \ldots, T}$.
We propose to use a similar relationship between $\tau_{lat}$, $\tau_{obs}$ and $c$, not only in the Gaussian case, but also for arbitrary bivariate copula families.
Therefore, in general, we impose the following identifiability constraint on the copula parameter for all bivariate copula families with a single parameter identified uniquely by Kendall's $\tau$ as follows
\begin{equation}
\sin\left(\frac{\pi}{2}\tau_{obs}\right) = \left(\sin\left(\frac{\pi}{2}\tau_{lat}\right) \right)^c \qquad \mbox{for some suitable value $ c \geq 1$.} \label{eq:idrel} 
\end{equation}

\section{Bayesian analysis of the copula state space model} \label{BayesAnalysis}
\subsection{Hamiltonian Monte Carlo}
The copula state space model is a highly non-linear and non-Gaussian model, which provides great flexibility by allowing for different bivariate copulas.
The downside of this flexibility is that inference for this model is not straight forward, e.g. it is not possible to implement a Gibbs sampler, where we can directly sample from the corresponding full conditionals.
For inference for the  copula state space model we rely on the No-U-Turn sampler of \cite{hoffman2014no} implemented within the STAN framework (\cite{carpenter2016stan}). The No-U-Turn sampler extends Hamiltonian Monte Carlo (HMC) and adaptively selects tuning parameters. HMC can be considered as a Metropolis Hastings algorithm, where new states are efficiently obtained by using information on the gradient of the log posterior density. The gradient is obtained through automatic differentiation (\cite{carpenter2015stan}) in STAN. The HMC sampler has shown good performance in several other cases (\cite{hajian2007efficient, pakman2014exact, hartmann2017bayesian}). We provide a short introduction to HMC in Appendix B and refer to \cite{neal2011mcmc} or \cite{betancourt2017conceptual} for more details.

An alternative Bayesian approach for jointly estimating parameters and states in non-linear non-Gaussian state space models is presented by \cite{barra2017joint}, who designed flexible proposal densities for the independent Metropolis-Hasting and the importance sampling algorithms.

\subsection{Posterior inference}
\label{sec:postinf}
As prior distribution for $\tau_{lat}$ we use a uniform prior on (0,1), which is a non-informative prior restricted to positive dependence, since we do not expect negative dependence in our application. With this prior choice we obtain a fully specified Bayesian model with posterior density
\begin{equation*}
\begin{split}
 \pi(\tau_{lat}, v_1, \ldots, v_T|\hat u_1, \ldots, \hat u_T) = \prod_{t=1}^T c_{U,V}(\hat u_t,v_t;\tau_{obs}) \prod_{t=2}^T c_{V_2,V_1}(v_t,v_{t-1};\tau_{lat}),
\end{split}
\end{equation*}
where 
$\tau_{obs}$ is a function of $\tau_{lat}$ as given in \eqref{eq:idrel}. Note that for the Bayesian approach the latent variables of the state equation are considered as parameters.
We run the No-U-Turn sampler to sample from this posterior density. For a chosen $c$ we obtain a posterior sample for $\tau_{lat}$ 
$$
\tau_{lat}^r (c), \hspace{0.5cm} r = 1, \ldots, R
$$
and, similarly, for $\tau_{obs}$, using the relationship in \eqref{eq:idrel},
$$
\tau_{obs}^r (c), \hspace{0.5cm} r = 1, \ldots, R
$$
where $R$ is the total number of HMC iterations. Additionally, posterior samples for the latent variables $V_t$, for $t= 1, \ldots, T$, are denoted by
$$
v_t^r (c), \hspace{0.5cm} t= 1, \ldots, T \hspace{0.5cm} \mbox{and} \hspace{0.5cm} r = 1, \ldots, R.
$$

\subsection{Predictive simulation}
\label{seq:pred}
An advantage of the Bayesian approach is that our model already specifies the predictive distribution, which is the distribution of the response for new data points conditional on observed data points. From this distribution uncertainty is easy to be quantified through credible intervals.

We consider a posterior sample of the model parameters given by the set \linebreak $\left\{ \tau_{lat}^r (c), \,\,\, v_t^r (c), \,\,\, r = 1, \ldots, R, \,\,\, t = 1, \ldots, T \right\}$. Simulations for a new value at time $t \in \{1, \ldots, T\}$ on the copula scale can be obtained by 
\begin{itemize}
\item simulate $u_{t}^r (c)$ from $\mathbbm{C}_{U|V}^{obs} \left(\cdot | v_{t}^r (c); \tau_{obs}^r (c)\right) $.
\end{itemize}

We refer to the corresponding distribution as the in-sample predictive distribution on the copula scale. The out-of-sample predictive distribution refers to new values at time $t>T$.
Simulated values from the one-day-ahead predictive distribution of $U_{T+1}$ given $U_T$, can be obtained as follows

\begin{itemize}
\medskip
\item simulate $v_{T+1}^r (c)$ from $\mathbbm{C}_{V_{2}|V_{1}}^{lat} \left(\cdot | v_T^r (c); \tau_{lat}^r (c)\right) $,
\medskip
\item simulate $u_{T+1}^r (c)$ from $\mathbbm{C}_{U|V}^{obs} \left(\cdot | v_{T+1}^r (c); \tau_{obs}^r (c)\right) $.
\end{itemize}
\medskip

In general, simulations from the $i$-days-ahead out-of-sample predictive distribution on the copula scale can be obtained recursively through:

\begin{itemize}
\medskip
\item simulate $v_{T+i}^r (c)  \sim \mathbbm{C}_{V_{2}|V_{1}}^{lat} \left(\cdot | v_{T+i-1}^r (c); \tau_{lat}^r (c)\right) $,
\medskip
\item simulate $u_{T+i}^r (c) \sim \mathbbm{C}_{U|V}^{obs} \left(\cdot | v_{T+i}^r (c); \tau_{obs}^r (c)\right) $.
\end{itemize}

Based on a simulation of the (in-sample or out-of-sample) predictive distribution  on the copula scale $u_{t}^r(c)$, we further define
$$
{\varepsilon}_{t}^r (c) := \Phi^{-1} \left( u_{t}^r (c) \right)
$$ 
as a sample of the predictive distribution of the error of the GAM model specified in \eqref{eq:GAM}. In particular we estimate $E(Y_t)$ by $\hat f(\boldsymbol x_t)$ with estimated error variance $\hat \sigma^2$. So,
$$
{y}_{t}^r (c) :=\hat{f}(\boldsymbol{x}_{t}) + \hat{\sigma} {\varepsilon}_{t}^r (c)
$$
gives a sample of the predictive distribution of the response.
Note that to obtain this predictive sample we ignore the uncertainty in the marginal distribution.

\section{Data analysis} \label{DataAnalysis}

Recall the hourly data set discussed in Section \ref{seq:data} divided into 12 sub data sets, one data set for each month. 

\subsection{Marginal models}
 For each of the 12 training data sets we fit a GAM using the R package $\texttt{mgcv}$ of \cite{wood2015package}, where the response is the logarithm of PM2.5 and the covariates are DEWP, TEMP, PRES, IWS, PREC and CBWD, as described in Section \ref{seq:data}.  We define an additional covariate PREC\_{ind}, which indicates if there is precipitation, i.e. PREC\_{ind} $= \mathbbm{1}_{\text{PREC}>0}$. We also use the hour denoted by H and the weekday denoted by D as covariates.
\cite{liang2015assessing} showed that the wind direction not only has influence on the response itself, but might also influence the relationship between the other covariates DEWP, TEMP, PRES, IWS, PREC and the response. Therefore we allow for different smooth terms corresponding to different wind directions. More precisely, we create four indicator variables corresponding to the four wind directions $\mathbbm{1}_{\text{CBWD=CV}}$, $\mathbbm{1}_{\text{CBWD=NE}}$, $\mathbbm{1}_{\text{CBWD=NW}}$ and $\mathbbm{1}_{\text{CBWD=SE}}$. Then, we replicate the part of the model matrix corresponding to a covariate $x$ four times and multiply each of the four parts with one of the indicator variables $\mathbbm{1}_{\text{CBWD=CV}}$, $\mathbbm{1}_{\text{CBWD=NE}}$, $\mathbbm{1}_{\text{CBWD=NW}}$ and $\mathbbm{1}_{\text{CBWD=SE}}$. So, we obtain four smooth terms for each of the covariates DEWP, TEMP, PRES and IWS.  We do not allow for these interactions with the covariate PREC since this variable has only few values not equal to zero. For variable selection the approach of \cite{marra2011practical} is used, which allows terms to be penalized to zero.

\begin{figure}[H]
\centerline{%
\includegraphics[trim={0 0cm 0 0},width=0.9\textwidth]{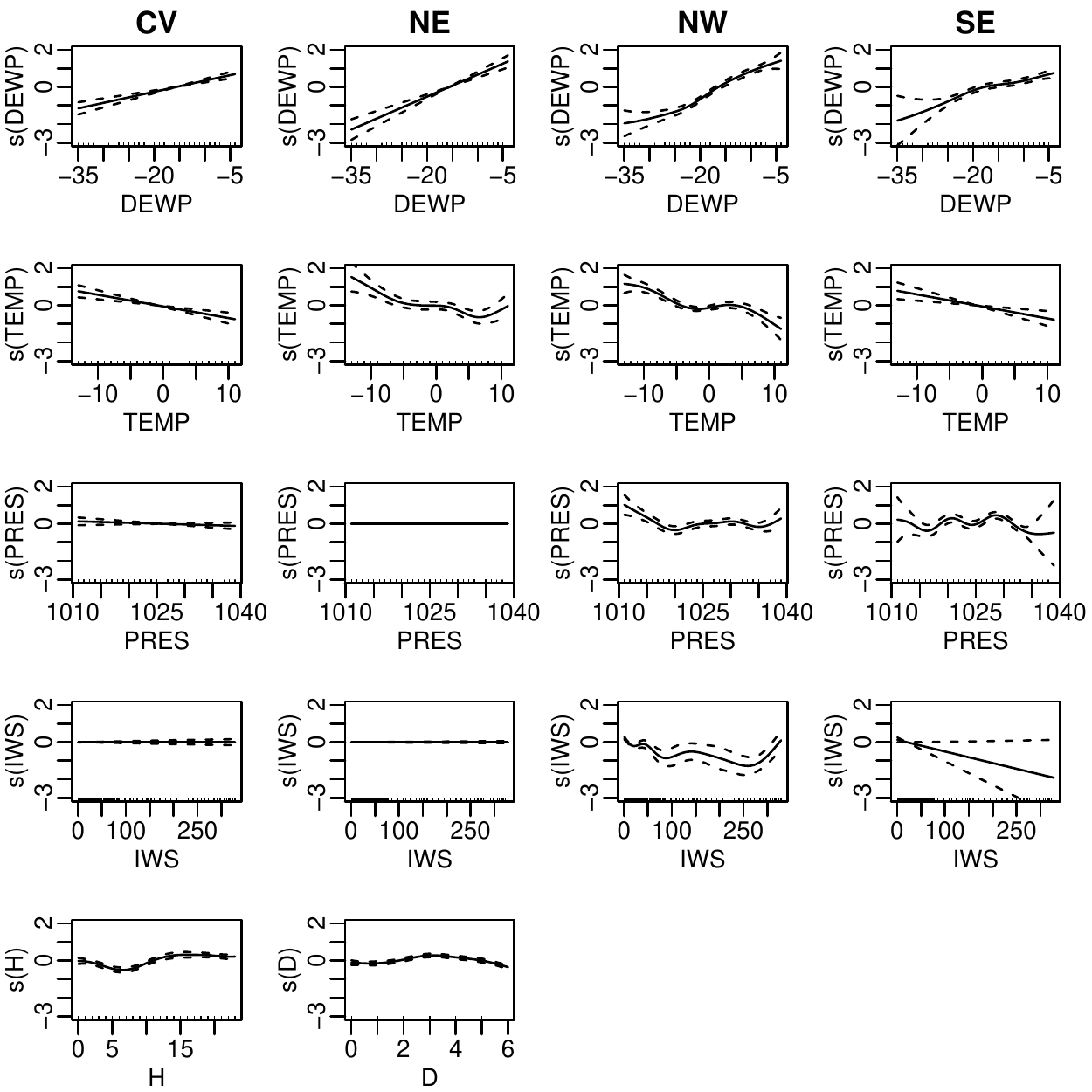}%
}%
\caption{Estimated smooth components of the GAM for month 1 for the covariates DEWP, TEMP, PRES, IWS, H and D. For each of the covariates DEWP, TEMP, PRES and IWS we have four different smooth terms corresponding to the four wind directions: CV, NE, NW, and SE. The dashed lines represent a pointwise $95\%$ confidence band.}
\label{fig:gam11}
\end{figure}

Plots of the different estimated smooth components are shown in Figure \ref{fig:gam11} for the January (month 1) data set. 
Plots of the estimated smooth terms in Figure \ref{fig:gam11} indicate the covariate effects on PM2.5. For example, with northwestern winds (NW), PM2.5 is lower for higher temperatures (TEMP). Furthermore, we draw the same conclusion as \cite{liang2015assessing}, that different smooth terms are necessary for different wind directions. For example, with northeastern winds (NE), we do not see any influence of the covariate PRES on PM2.5, whereas with northwestern winds (NW), we observe some non linear relationship between PRES and PM2.5.

\subsection{Model selection of monthly  copula family and value of c based on the Watanabe Akaike Information Criterion (in-sample)}
We now consider model selection for the copula state space model. This includes the selection of the copula families and the selection of the value of $c$. We fit models with different copula families and different values of $c$ and select the model which minimizes the Watanabe Akaike Information Criterion (WAIC) introduced by \cite{watanabe2010asymptotic}. For our model AIC and BIC would require to integrate out all the latent variables. Therefore we stick to the WAIC which is easy to evaluate for such Bayesian models with latent variables. We define by $\ell_t^r := c(\hat u_t,v_t^r;\tau_{obs}^r(c))$ the likelihood contribution of iteration $r$ at time $t$. Following \cite{vehtari2017practical} the WAIC can then be estimated by
\begin{equation*}
WAIC = -2\sum_{t=1}^T \left[ \ln\left( \hat {\text{E}}( (\ell_t^r)_{r=1, \ldots, R})\right) - \widehat{\text{ Var}}\left( (\ln(\ell_t^r))_{r=1, \ldots, R}  \right)\right],
\end{equation*}
where $\hat {\text{E}}$ denotes the sample mean and $ \widehat{\text{Var}}$ the sample variance.

We have one GAM specification for each month and obtain, for each month, approximately Uniform(0,1) pseudo-copula data $\hat u_{t}$ by the probability integral transform $\hat u_{t} = \Phi \left( \frac{y_{t}-\hat f(\boldsymbol x_{t})}{ \hat\sigma} \right)$ for $t=1, \ldots, T$ as in \eqref{eq:uresidual}. Here $\hat f$ and $\hat \sigma$ are the estimates of the GAM and $T$ denotes the number of observations in the corresponding monthly data set. To simplify notation we avoid indexing the models by month.

In the following we study several models that can be divided into three model classes.

\begin{itemize}
\item \textbf{Gaussian state space model $\mathscr{M}_{Gauss}$}: 
$\mathbbm{C}^{obs}_{U,V}$ and $\mathbbm{C}^{lat}_{V_2,V_1}$ are both Gaussian copulas.
\item \textbf{Copula based state space model $\mathscr{M}_{Cop}$}:
$\mathbbm{C}^{obs}_{U,V}$ and $\mathbbm{C}^{lat}_{V_2,V_1}$ are from the same bivariate copula family.
\item \textbf{GAM model with independent errors $\mathscr{M}_{Ind}$}: 
$\mathbbm{C}^{obs}_{U,V}$ and $\mathbbm{C}^{lat}_{V_2,V_1}$ are both independence copulas.
This corresponds to a standard GAM model with independent errors.
\end{itemize}

For each of the 12 monthly data sets on the copula scale, the three model classes are fitted. To estimate model parameters we run the No-U-Turn sampler with 2 chains, where each chain contains 2000 iterations. The first 500 iterations are discarded for burnin. Preliminary analysis showed that this burnin choice is sufficient.
We fit the independence model $\mathscr{M}_{Ind}$, the Gaussian model $\mathscr{M}_{Gauss}$ for every value of $c=1,3,6,10$ and several latent copula models for the class $\mathscr{M}_{Cop}$.
The different state space copula models correspond to all combinations of the values of $c = 1,3,6,10$ and the following bivariate parametric copula families: Student t (df=3), Student t (df=6), Gumbel, Clayton and Frank. This set includes copula families that are appropriate for the observed contour plots in Figure \ref{fig:contour}.
So for one specific monthly data set a model is specified by the value of $c$ and the copula family.

\begin{figure}[H]
\centerline{%
\includegraphics[trim={0 9.5cm 0 0},width=1\textwidth]{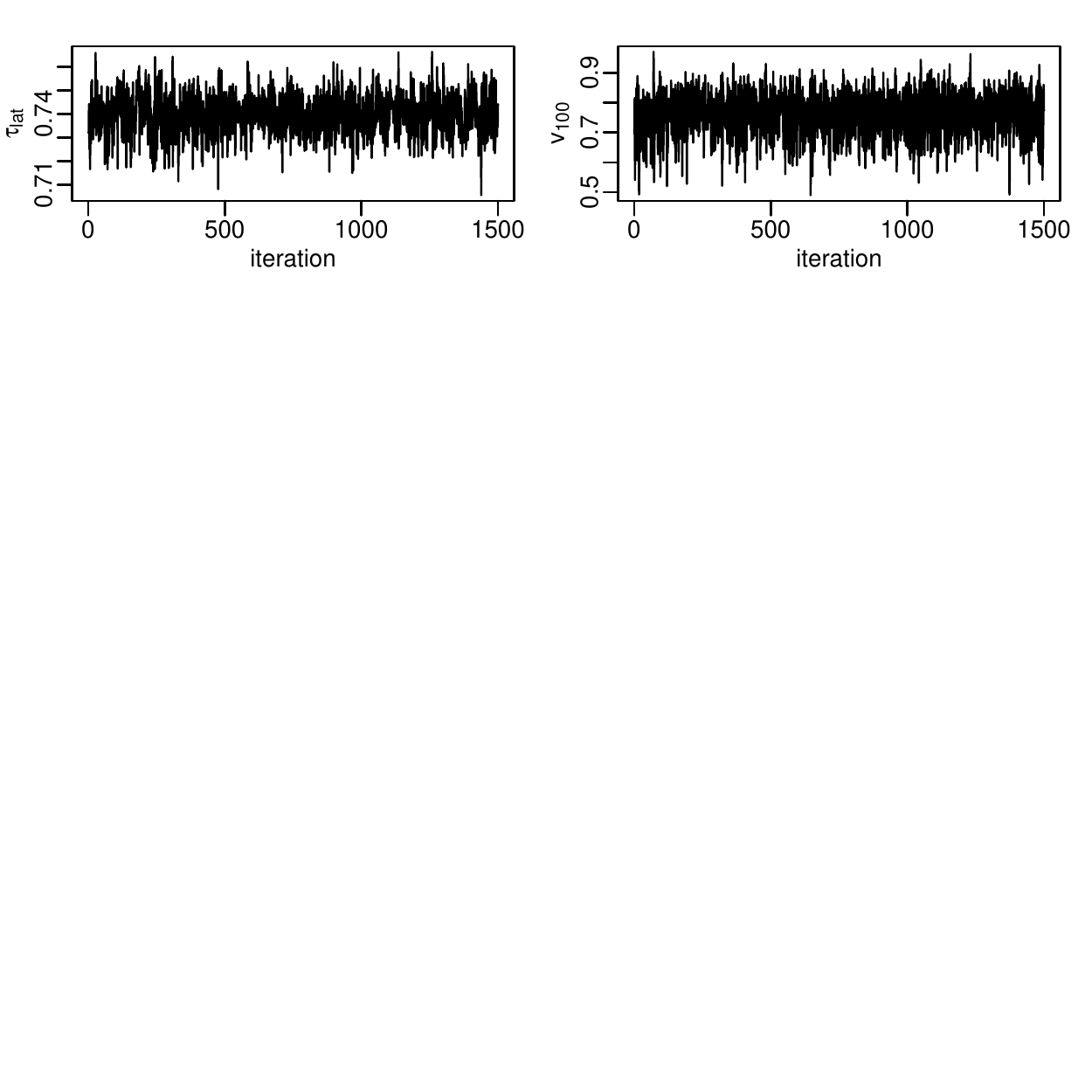}%
}%
\caption{Traceplots of 1500 posterior draws after a burnin of 500 iterations of $\tau_{lat}$ (left) and $V_{100}$ (right) of the first chain of the HMC sampler for the model with Student t copulas with 6 degrees of freedom and $c=1$ using the data set for January.}
\label{fig:tplot_onerun}
\end{figure}

\begin{figure}[H]
\centerline{%
\includegraphics[trim={0 6.25cm 0 0},width=0.9\textwidth]{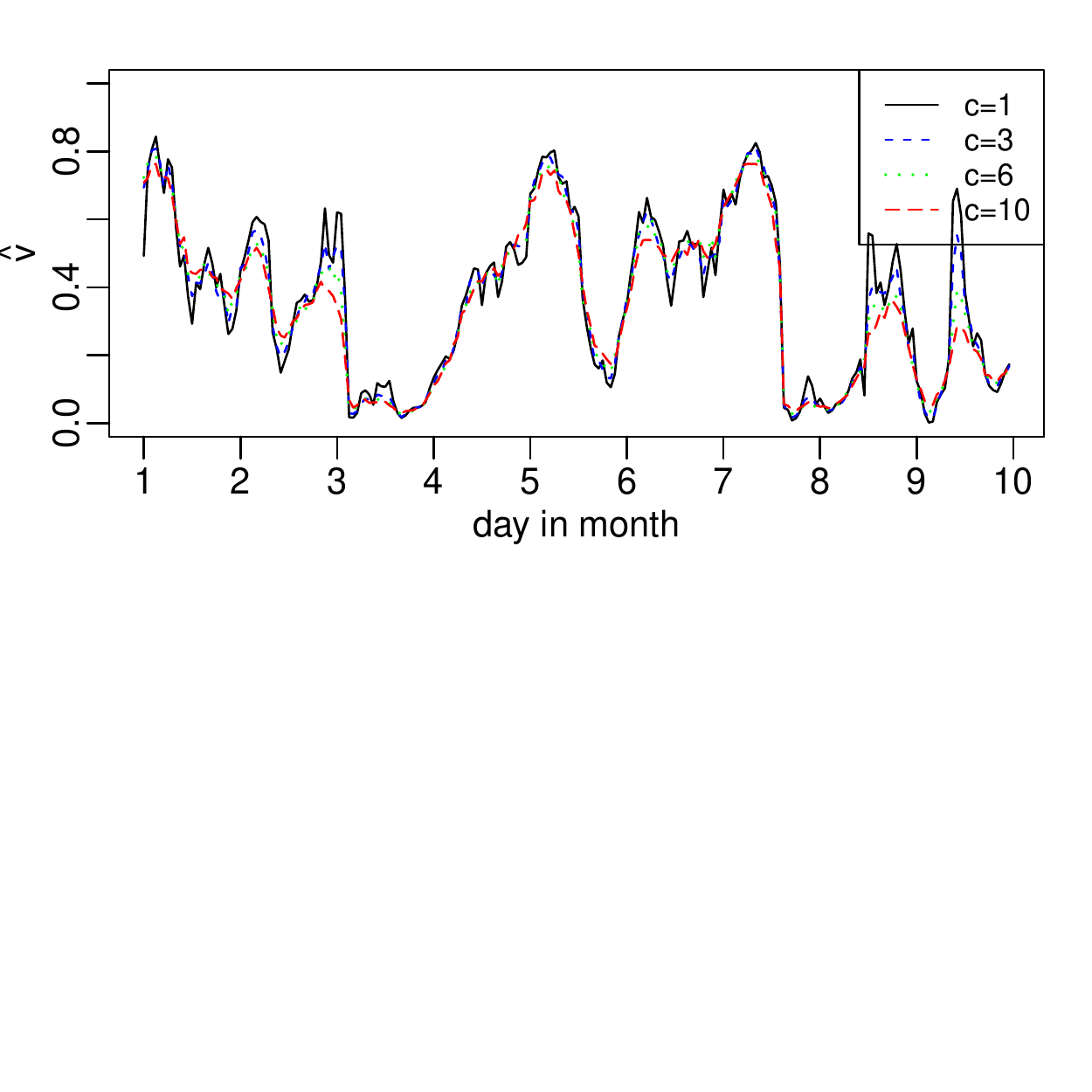}%
}%
\caption{Estimated hourly posterior mode of the latent state $\hat{v_t}$ at time $t$ plotted against $t$ for the first 9 days of January for models with Student t copulas with 6 degrees of freedom and different values of $c$ ($c=1,3,6,10$). The posterior mode estimates are obtained as modes of univariate kernel density estimates and are based on 3000 iterations from two chains.}
\label{fig:fit_expo}
\end{figure}

As an example we have a closer look at the model for January with Student t copulas with 6 degrees of freedom and $c=1$. Figure \ref{fig:tplot_onerun} shows the traceplots of the dependence parameter $\tau_{lat}$ and the latent state at time point $100$ ($V_{100}$) for the first chain. The traceplots suggest that the chains have converged. The chain for $\tau_{lat}$ converges to values far away from zero, thus showing dependence. Figure \ref{fig:fit_expo} illustrates the effect of the different values of $c$ on the posterior mode estimates of the latent states $\hat v_t$. As expected, we observe that the size of the oscillations decrease as the value of $c$ increases. 

Table \ref{tab:best_mod} shows the best model in $\mathscr{M}_{Cop}$, characterized by the value of $c$ and the copula family, and the best model in $\mathscr{M}_{Gauss}$, characterized by the value of $c$. In addition Table \ref{tab:best_mod} shows the WAIC of the best model within the model classes $\mathscr{M}_{Cop}$, $\mathscr{M}_{Gauss}$ and $\mathscr{M}_{Ind}$. We see that for $\mathscr{M}_{Gauss}$ and $\mathscr{M}_{Cop}$ the value of $c$ of the best model is always equal to 1 thus allowing for higher oscillations in the posterior of the latent states.  
The best model according to the WAIC is provided by the copula based model class $\mathscr{M}_{Cop}$ for every month, since this model is always associated to the smallest WAIC.

\begin{table}[H]
\centering
\caption{Family of the best model in $\mathscr{M}_{Cop}$, value of c of the best model in  $\mathscr{M}_{Cop}$ and the best model in  $\mathscr{M}_{Gauss}$ and the WAIC of the best model within each class $\mathscr{M}_{Cop}$, $\mathscr{M}_{Gauss}$ and $\mathscr{M}_{Ind}$. The best model is selected with respect to the WAIC.}
\begin{tabular}{l|l|ll|rrr}
& family &  \multicolumn{2}{c}{$c$} & \multicolumn{3}{c}{$WAIC$} \\
month&$\mathscr{M}_{Cop}$&$\mathscr{M}_{Cop}$&$\mathscr{M}_{Gauss}$&$\mathscr{M}_{Cop}$&$\mathscr{M}_{Gauss}$&$\mathscr{M}_{Ind}$ \\
\hline
1 & t(6) & 1 & 1 & -926 & -887 & 0 \\ 
  2 & Frank & 1 & 1 & -755 & -702 & 0 \\ 
  3 & Frank & 1 & 1 & -1000 & -898 & 0 \\ 
  4 & t(3) & 1 & 1 & -1200 & -1103 & 0 \\ 
  5 & t(6)  & 1 & 1 & -982 & -945 & 0 \\ 
  6 & t(3) & 1 & 1 & -672 & -604 & 0 \\ 
  7 & t(3)  & 1 & 1 & -808 & -722 & 0 \\ 
  8 & t(3) & 1 & 1 & -680 & -653 & 0 \\ 
  9 & t(6) & 1 & 1 & -972 & -873 & 0 \\ 
  10 & Gumbel & 1 & 1 & -1130 & -1102 & 0 \\ 
  11 & t(6) & 1 & 1 & -910 & -900 & 0 \\ 
  12 & t(6) & 1 & 1 & -765 & -758 & 0 \\ 
\end{tabular}

\label{tab:best_mod}
\end{table}

\subsection{Analysis of fitted models}\label{InSample}

\begin{figure}[H]
\centerline{%
\includegraphics[trim={0 3cm 0 0},width=1.0\textwidth]{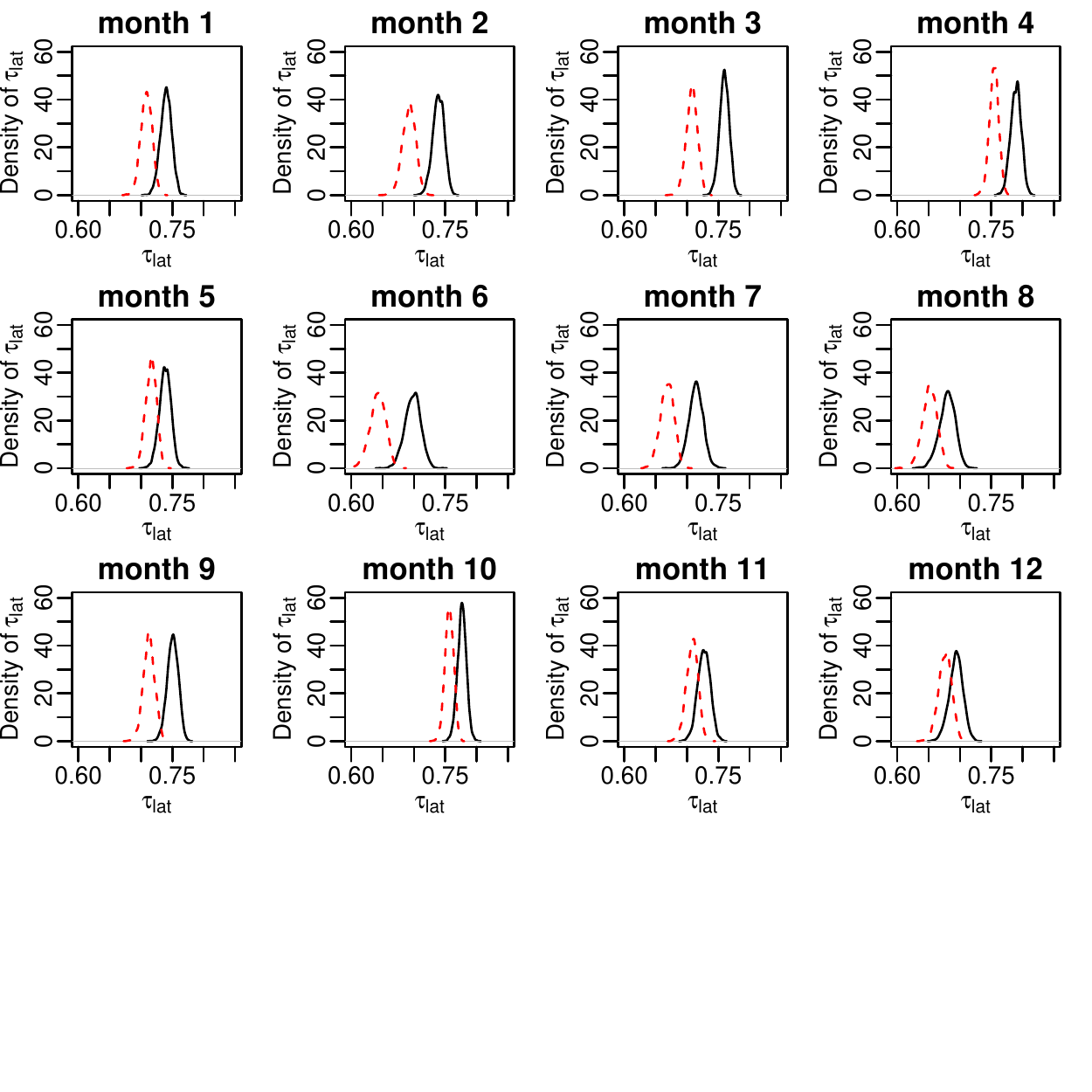}%
}%
\caption{Estimated posterior density of the dependence parameter $\tau_{lat}$ for the best model in $\mathscr{M}_{Cop}$ (black) and $\mathscr{M}_{Gauss}$ (red, dashed) according to the WAIC for all 12 data sets.}
\label{fig:taudens_best}
\end{figure}

In the previous section we selected the best copula state space models according to the lowest WAIC. This gave the copula family choice and the value of $c$ for $\mathscr{M}_{Cop}$ and the value of $c$ for $\mathscr{M}_{Gauss}$. 
Figure \ref{fig:taudens_best} shows the estimated posterior densities for the dependence parameter $\tau_{lat}$ for these models. We observe that most of the mass of the posterior density concentrates between $0.6$ and $0.8$ for all monthly models. This range for $\tau_{lat}$ coincides with positive dependence between two succeeding time points.
We also see that the Kendall's $\tau$ values of the $\mathscr{M}_{Cop}$ model class are slightly higher than those of the $\mathscr{M}_{Gauss}$ model class for all months.

\begin{figure}[H]
\centerline{%
\includegraphics[trim={0 7cm 0 0},width=1.0\textwidth]{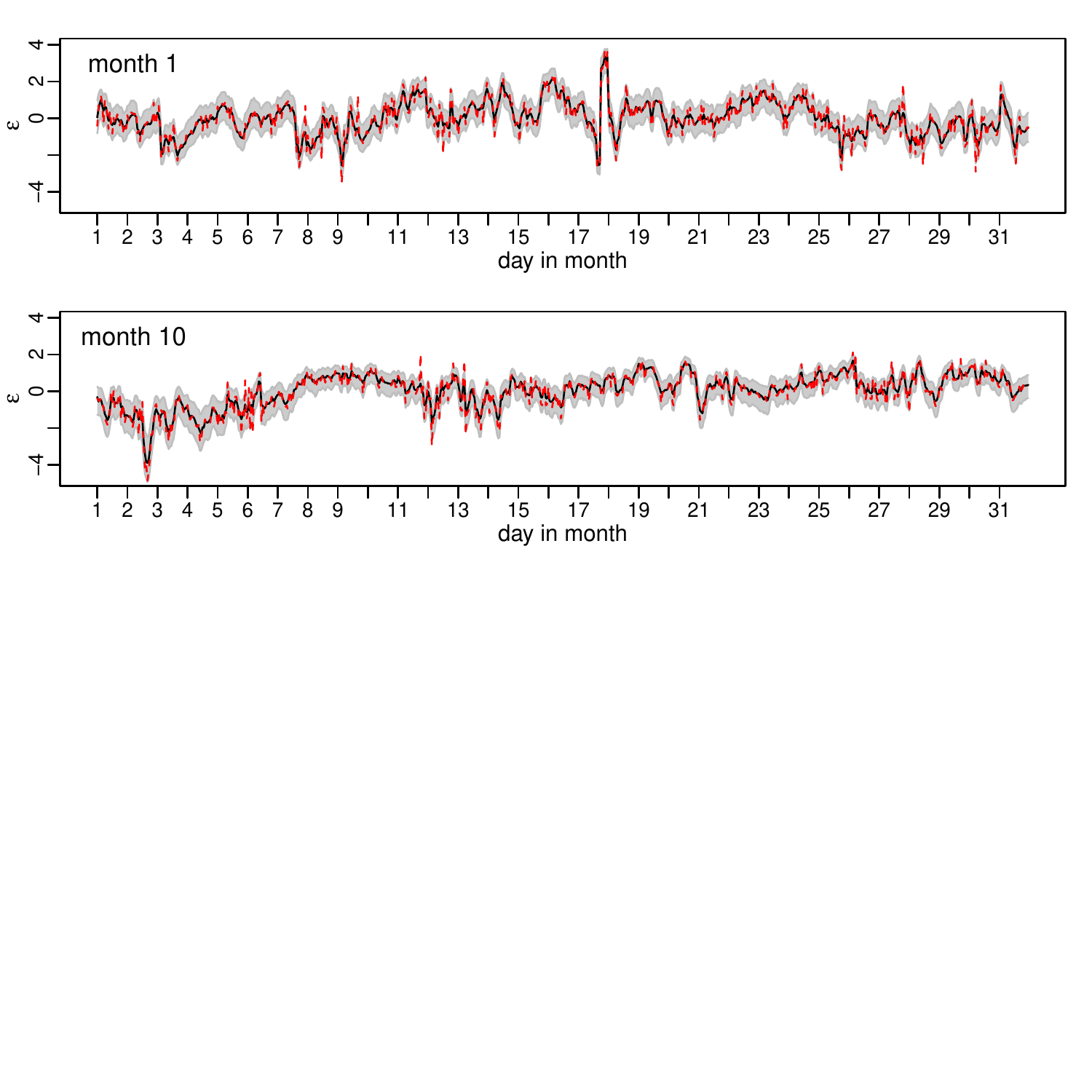}%
}%
\caption{Estimated mode of the predictive density of the error $\epsilon_t$ plotted against $t$ for every data point in January (top row) and October (bottom row) using the best models in $\mathscr{M}_{Cop}$ as selected by WAIC. A $90\%$ credible region, constructed from the $5\%$ and $95\%$ empirical quantiles of simulations from the predictive distribution of the error, is added in grey. Further, the standardized residual of the GAM $\hat z_t$ is added in red (dashed).}
\label{fig:simZ}
\end{figure}

The copula based state space model was fitted to the standardized residuals of the GAM $\hat z_t$ as defined in \eqref{eq:hatresiduals}.
To further evaluate our model, we simulate from the predictive distribution of the error for each $t \in \{1, \ldots, T\}$, as explained in Section \ref{seq:pred}, and compare it to the standardized residuals of the corresponding GAM model. Figure \ref{fig:simZ} shows that the copula based state space model is able to recover the dynamics of the standardized residuals.

If we ignored the latent effect, the distribution of the error would be standard normal. Simulating from the predictive distribution of the error can be considered as taking the latent effect into account. Therefore a concentration of the predictive distribution that is far away from zero indicates time points where the latent variable has higher effects. These are time points where the level of the response is unusually high or low for the corresponding specification of the covariates.

We see from Figure \ref{fig:simZ} that on January 18th, the estimated mode of the predictive density of the error is high. On this day unusual high pollution was recorded in Beijing where PM2.5 reached around 500 micrograms per cubic meter ($\mu$g/$\mbox{m}^3$), skyrocketing to more than 20 times the level considered unhealthy by the World Health Organization
   \footnote{See \url{http://www.takepart.com/article/2014/01/18/beijing-china-air-pollution-billboard} }. The copula based state space model with a Student t copula has a high peak on that day and is able to capture this unusual behaviour.

The ability to model unusual high peaks of airborne contaminants is fundamental to accurately assess the effect of exposure to human health.
Indeed, many studies in the literature show that increased levels of air pollutants may have a dramatic effect on human health.
In particular, for a 10-$\mu$g/$\mbox{m}^3$ increase in PM2.5, hospital admissions for ischemic cardiac events and heart failures may increase by $4.5\%$ and $3.6\%$, respectively; respiratory and pneumonia hospitalizations may increase by $17\%$ and $6.5\%$; respiratory and lung cancer mortality may increase by $2.2\%$ and $8\%$. In addition, exposure to PM2.5 is estimated to reduce the life expectancy of the population by about $8.6$ months on average (\cite{anderson2012clearing}).
The economic impact of PM2.5 pollution is also relevant, since fine particulate matter-related illness can ultimately lead to financial and non-financial welfare losses of not only patients and their families but also a significant portion of gross domestic product (GDP). Indeed, it was estimated that in 2009 China suffered a health-related economic loss of $2.1\%$ of its GDP, corresponding to $106.5$ billion US dollars (\cite{kim2015review}).
Therefore, the consequences on citizens' health and economy of an extremely high value of PM2.5, such as the one expereienced in Beijing on the 18th January 2014, may be very severe and extensive.

The proposed Bayesian non-linear non-Gaussian state space model allows us to capture unusual extreme air pollution events appropriately and could provide accurate information to stakeholders such as doctors and policy makers to better evaluate the consequences of pollution on citizens.

\subsection{Out-of-sample predictions}
Short term predictions of PM2.5 levels can be used to alert citizens of high pollution periods which are dangerous to health. In this section we construct predictions several hours up to two days ahead. More precisely, we consider the best copula state space  model for March and use it to predict the first 48 hours of April. We choose March, since it is the month for which the non-elliptical Frank copula was selected.

\begin{figure}[H]
\centerline{%
\includegraphics[trim={0 0cm 0 0},width=0.6\textwidth]{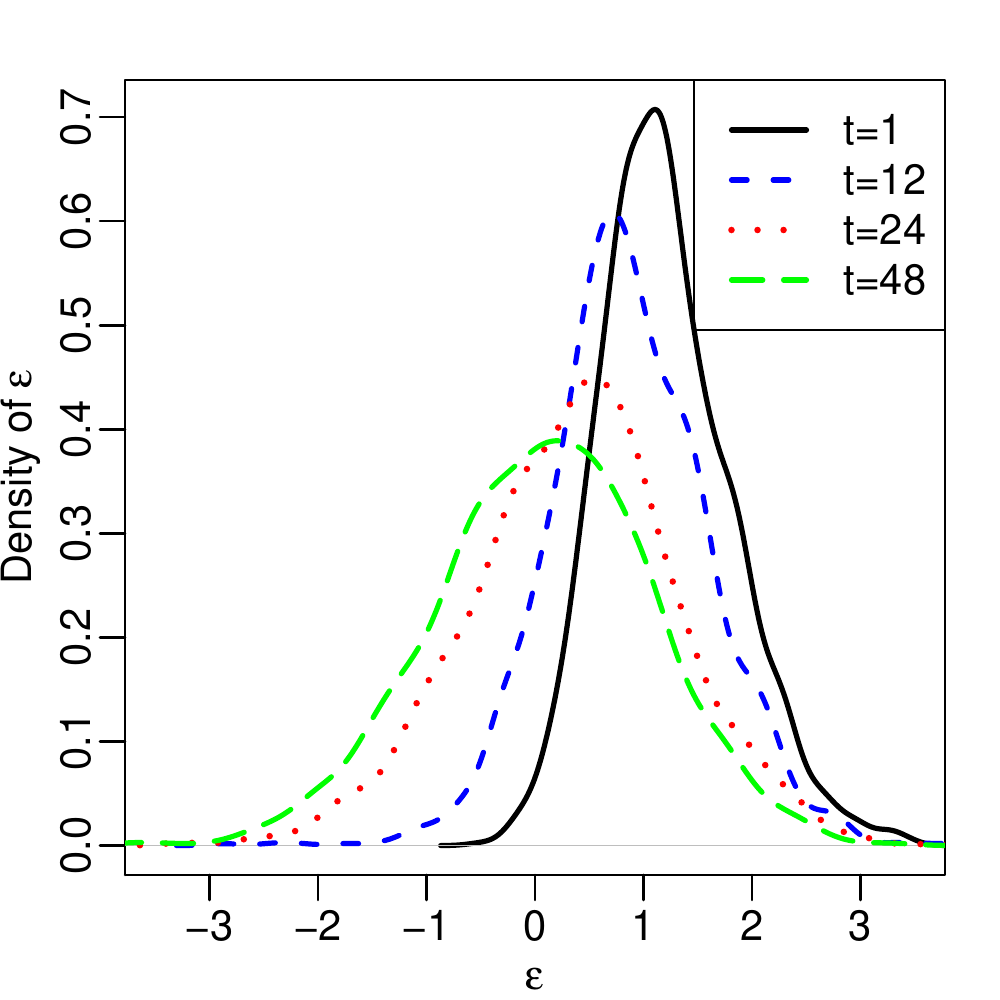}%
}%
\caption{Estimated predictive density of $\epsilon_{T+t}$ using the best copula state space model for March for different time steps (hours) ahead ($t=1,12,24,48$). The estimated predictive density is the kernel density estimate of simulations from the corresponding predictive distribution.} 
\label{fig:pdens_onerun}
\end{figure}

We first simulate from the out-of-sample predictive distribution of the error as explained in Section \ref{seq:pred}. Figure \ref{fig:pdens_onerun} shows predictive densities for different time-steps ahead for this model, more precisely the estimated forecast density of $\epsilon_{T+t}$ for $t=1, 12, 24, 48$ hours based on 3000 HMC iterations from two chains. As we see from Figure \ref{fig:pdens_onerun}, we obtain non-Gaussian forecast densities. Further, the densities are more disperse for a longer time period ahead, reflecting the fact that uncertainty increases if we predict a longer time period ahead.

\begin{figure}[H]
\centerline{%
\includegraphics[trim={0 6.5cm 0 0},width=0.85\textwidth]{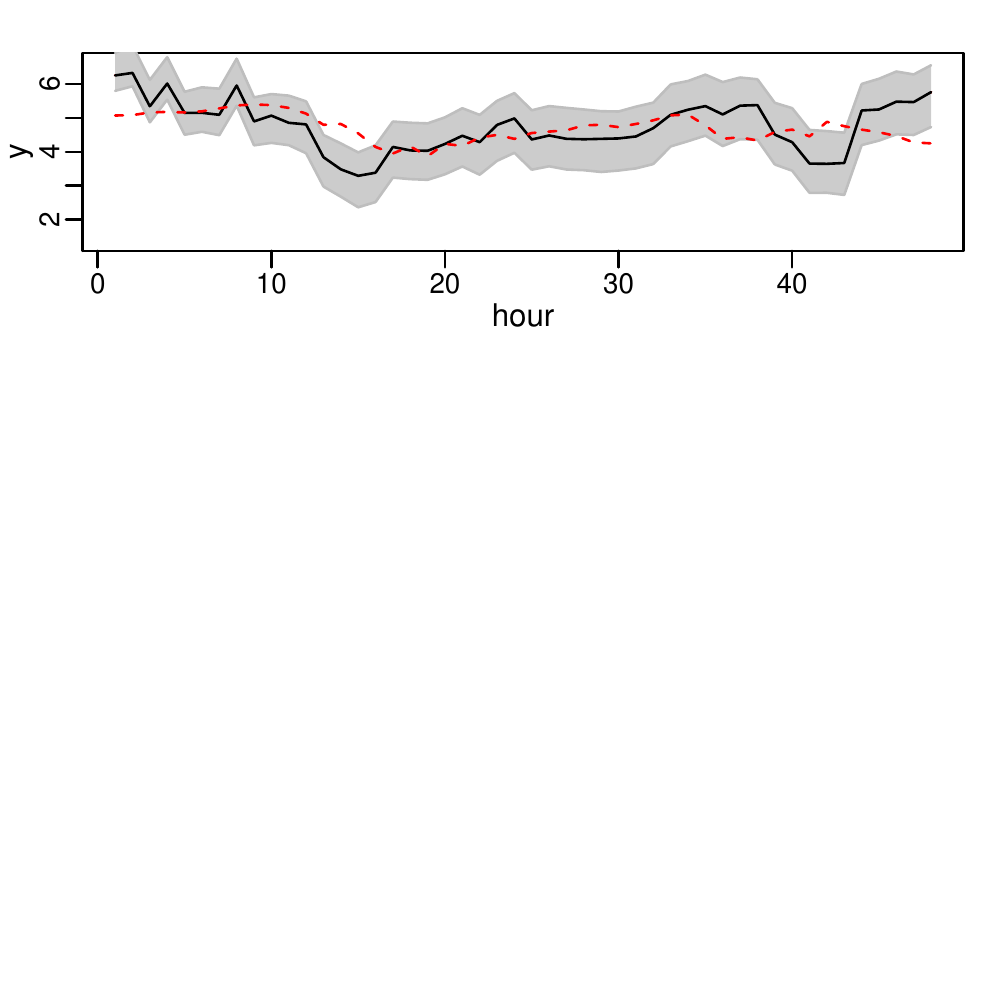}%
}%
\caption{
Estimated mode of the predictive density of the response  $t$ hours ahead plotted against $t$.  A $90\%$ credible region, constructed from the $5\%$ and $95\%$ empirical quantiles of simulations from the predictive distribution of the response, is added in grey. Further, the observed response values are added in red (dashed).
The simulations of the predictive distribution of the response 1 up to 48 hours ahead are obtained according to \eqref{eq:preddist_resp} based on the best copula state space model for March.} 
\label{fig:pred_feb}
\end{figure}

To obtain predictions for the PM2.5 levels the simulations for the error needs to be combined with the mean prediction of the GAM, according to our model 
\begin{equation*} 
Y_t = f(\boldsymbol{x}_t) + \sigma \varepsilon_t.
\end{equation*} 
To obtain the predicted mean of the GAM the covariate values are required. Except for the weekday D and the hour H, future covariate levels are not known. As a proxy for an unknown covariate vector with hour H$=$h, we use the covariate specifications of the last observed time point with the same hour H$=$h. 
We denote this covariate vector by $\boldsymbol{x}_{t}^l$ and obtain predictive simulations of the response at time $t>T$ as follows

\begin{equation} 
y_t^r = \hat f(\boldsymbol{x}_{t}^l) + \hat\sigma \varepsilon_t^r,
\label{eq:preddist_resp}
\end{equation}
for $r=1, \ldots, R$. These predictive simulations are visualized in Figure \ref{fig:pred_feb}. We see that the observed values are most of the time within the $90\%$ credible interval.

\begin{figure}[H]
\centerline{%
\includegraphics[trim={0 3cm 0 0},width=0.85\textwidth]{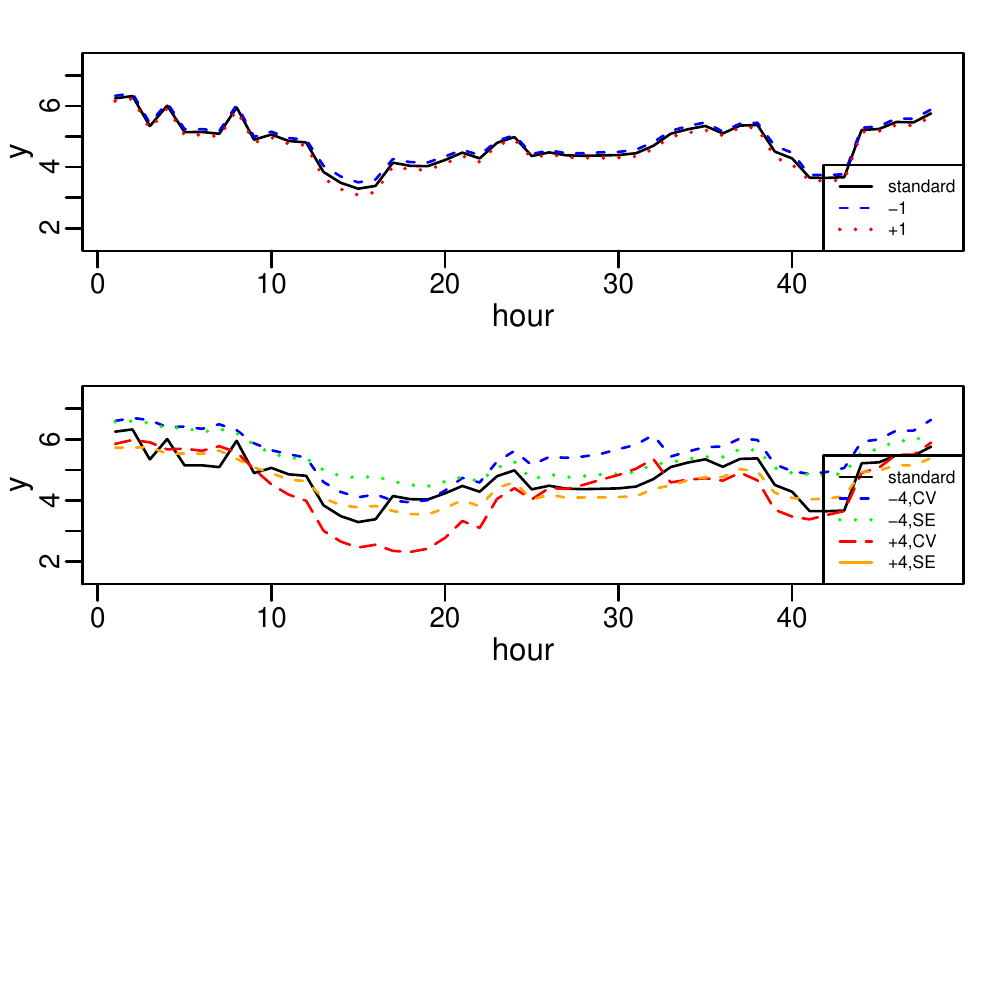}%
}%
\caption{
We show the estimated mode of the predictive density of the response  $t$ hours ahead plotted against $t$ for different specifications of the covariates. 
The simulations of the corresponding predictive distribution of the response 1 up to 48 hours ahead are obtained according to \eqref{eq:preddist_resp} based on the best copula state space model for March (black line). 
In the top row, we consider additionally predictive distributions where the temperature of $\boldsymbol{x}_{t}^l$ is changed by $\pm 1$ degree. In the bottom row, we consider additionally predictive distributions where the temperature of $\boldsymbol{x}_{t}^l$ is changed by $\pm 4$ degree and the covariate CBWD is set equal to SE or CV.
}
\label{fig:pred_feb_scen}
\end{figure}

In addition, the simulations for the error may be combined with mean predictions obtained from the GAM with different covariate specifications. Since the covariates several hours ahead are random, different scenarios as specified by different covariate levels are possible and should be taken into account. Here, we first consider two cases where the temperature at each time point in $\boldsymbol{x}_{t}^l$ is increased and decreased by 1 degree. Second, we also investigate more extreme scenarios for $\boldsymbol{x}_{t}^l$ where we decrease and increase the temperature at each time point by 4 degrees and in addition change the wind direction at each time point to the same value. The value for the wind direction CBWD is set to either CV or SE. This yields four different scenarios. The mode estimates of the resulting predictive densities are visualized in Figure \ref{fig:pred_feb_scen}. It is not surprising that the first case where we only change the temperature by 1 degree results in less changes in the mode estimates compared to the more extreme case. There are many more scenarios that can be analysed in a similar fashion. In particular, relevant scenarios suggested by experts could be analysed. A conservative warning system could alert citizens if at least one of the scenarios results in dangerous air pollution levels.

\subsection{Simulated scenarios}

Instead of only considering predictions several hours or days ahead, our model allows us to simulate typical air pollution levels that might occur in the same month in another year with different covariate levels.

 We may consider a different covariate vector $\boldsymbol{x}_t^{new}$ and obtain  
\begin{equation} 
(PM_t^{new})^r = \exp(\hat f(\boldsymbol{x}_t^{new}) + \hat \sigma \varepsilon_t^r),
\label{eq:pred_dist_lterm}
\end{equation} 
for $r = 1, \ldots, R$, where $\hat f$ and $\hat \sigma$ are estimates from the marginal GAM models and $\varepsilon_t^r$ is a simulation from the in-sample predictive distribution of the error, based on the data for 2014.
The values of $(PM_t^{new})^r$ give rise to typical air pollution levels that might occur in the same month in another year with covariate levels $\boldsymbol{x}_t^{new}$.

\begin{figure}[H]
\centerline{%
\includegraphics[trim={0 4cm 0 0},width=0.85\textwidth]{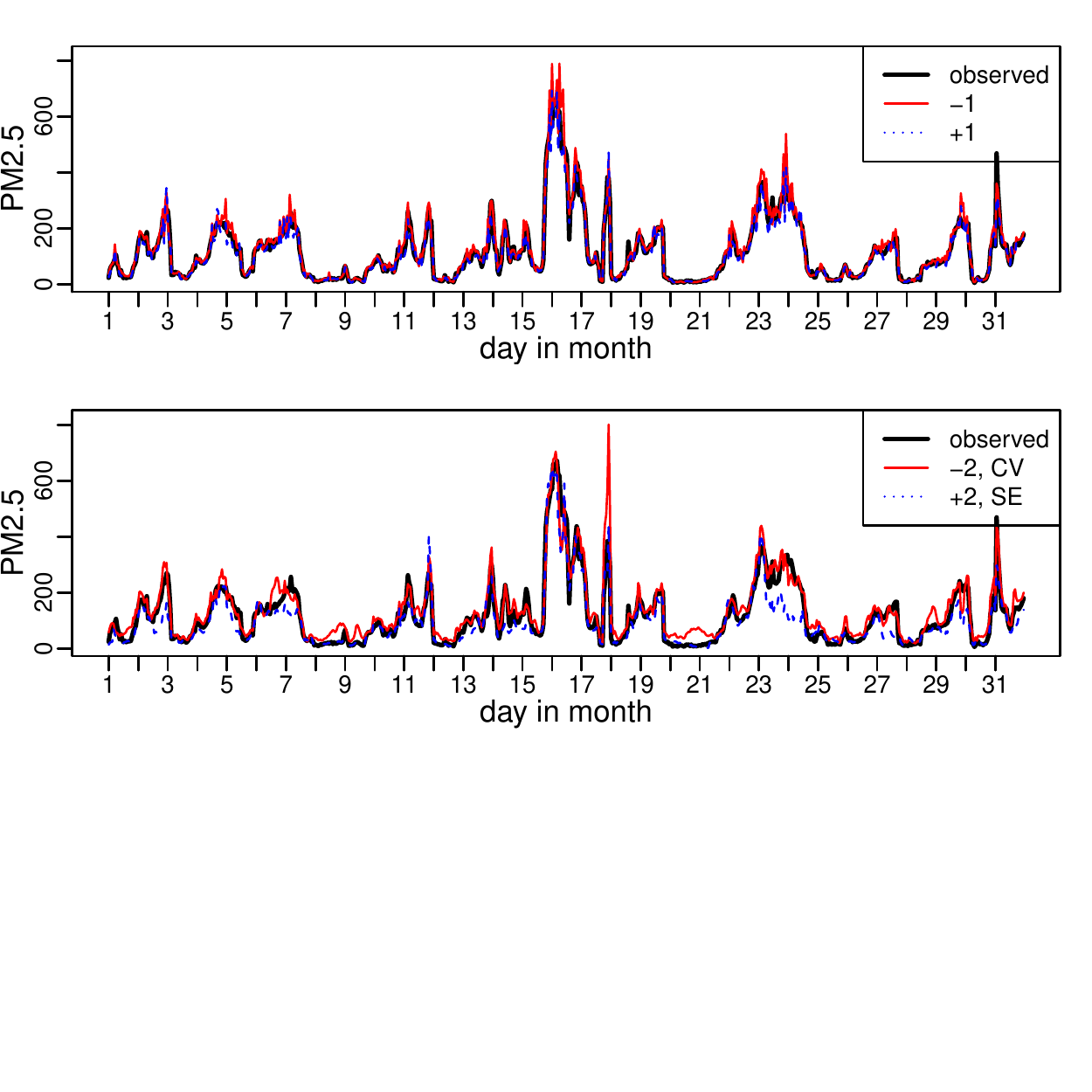}%
}%
\caption{
We show typical PM2.5 levels for January for different specifications of the covariates. The typical PM2.5 level is estimated as the mode of the kernel density estimate of simulations obtained as explained in \eqref{eq:pred_dist_lterm}. The top row shows typical air pollution levels where the temperature was changed by $\pm 1$ degree. In the bottom row we show one case where the temperature was decreased by 2 degree and the covariate CBWD was set equal to CV and another case where the temperature was increased by 2 degree and the covariate CBWD was set equal to SE. The other covariates are kept at the same levels as they were observed in 2014. The PM2.5 level observed in 2014 is added in black.} 
\label{fig:pdens_scens_lterm}
\end{figure}

Here we analyse different scenarios for January.
First we consider scenarios where we only change the temperature, leaving all the other covariates as they are.
We consider one case where we increase the original temperature variable at each time point by 1 degree and one case where it is decreased by 1 degree.
From the data set analysed by \cite{liang2015assessing}, of which our data set is a subset, we can see that differences of about 1 degree in the monthly average temperature between two different years are common. Second we investigate more extreme scenarios where we shift temperatures by $\pm 2$ degree and also change the wind direction. The dominant wind direction in January 2014 was NW (northwestern). The wind direction CBWD at each time point is now changed to the same value. The value is set equal to CV (calm and variable), NE (northeastern) or SE (southeastern).  Combining these three choices for the wind direction with two different choices for the temperature leads to 6 different scenarios.

In Figure \ref{fig:pdens_scens_lterm} we compare the mode estimates of the density of $\text{PM}_t^{new}$  to the observed PM2.5 values in 2014. We see that in January a decrease in temperature by one degree leads to higher pollution levels. We obtain higher peaks and the average PM2.5 level of this month increases  from 118 $\mu$g$/\mbox{m}^3$, as observed in January 2014,  to 127 $\mu$g$/\mbox{m}^3$. 
Further, we show in Figure \ref{fig:pdens_scens_lterm} the two out of the six more extreme cases that lead to the largest increase and decrease in the average PM2.5 level. Increasing the temperature by 2 degree and setting the wind direction equal to SE, leads to the largest decrease in the PM2.5 level. The average PM2.5 level decreases from 118 $\mu$g$/\mbox{m}^3$ to 95 $\mu$g$/\mbox{m}^3$. By decreasing the temperature by 2 degrees and setting the wind direction equal to CV (calm and variable), the average PM2.5 level increases from 118 $\mu$g$/\mbox{m}^3$ to 137 $\mu$g$/\mbox{m}^3$. Further, this scenario leads to higher peaks of the air pollution level. Our analysis shows that it is not unlikely to observe higher air pollution levels in future Januaries compared to those of January 2014.


\section{Summary and Outlook}\label{Conclusions}

The starting point of this paper was the question of how to capture not only non-linear effects of meteorological variables on  pollution measures such as airborne particulate matter, but also to allow for further time dynamics of the observations not covered by the meteorological variables. For this we investigated hourly data of ambient air pollution in Beijing and illustrated that the lag-one time dynamics is not a Gaussian one, thus ruling out standard linear state space models.

To deal with this non-Gaussian dependence we proposed a novel non-linear state space model based on a copula formulation for univariate observation and state equations. The observation and state variables are coupled using two bivariate copulas. Since the copula approach allows for separate modeling of the margins and dependence, the observation variables are allowed to follow any time invariant statistical model. In the application we utilized a GAM to allow for non-linear effects of covariates. Once the marginal distribution of the response variables is specified, they can be transformed to the uniform scale using the probability integral transform. The resulting value on the uniform scale at time $t$, $U_t$, is then coupled with a [0,1] valued state variable for time $t$ using a bivariate copula. Therefore, the observation equation of the copula based state space formulation is given by the conditional distribution of $U_t$ given the value of the state variable at time $t$.  The time dynamics of the state variables is then similarly 
modeled as the conditional distribution of the state variable at time $t$ given the state variable at time $t-1$, where these two state variables are jointly modeled by a bivariate copula. We first show that, in the case of bivariate Gaussian copula, standard linear state space models result. Since many different parametric bivariate copulas exist, the flexibility of the copula-based state space model is evident and thus a significant extension of linear Gaussian state space models is possible.

Of course, such an extension has its price. In our case this means we cannot follow a standard estimation 
approach as provided by the Kalman filter for linear state space models. Therefore we propose and develop a Bayesian approach based on HMC. Further we deal with some identifiability issues of the copula state space, which we solve by restricting the strength of the dependence among the lag-one state space variables to be at least as high as the one of the observation variable $U_t$ and the state variable at time $t$. 

The state variables can be interpreted as a way to capture non-measured effects and thus are very appropriate for the data set analyzed in this paper. It allowed us to identify unusual high levels of pollution, which were not captured by the measured variables. We also present, with appropriate normalized bivariate contour plots, explorative tools to detect non-Gaussian dependence structures. 

The proposed approach can be used to accurately model extreme air pollution events and can assist stakeholders in the evaluation of the health consequences of exposure.
The analysis of high temporal resolution particulate matter data allows us to immediately detect quick upsurges of airborne contaminants and anticipate lower temporal resolution health effects.
The ability to predict future levels of fine particulate matter is another feature of our model, that was used to simulate different scenarios, in absence of future values of the covariates. Stakeholders may benefit from the prediction of PM2.5 levels associated to different meteorological conditions. 

The approach first proposed here allows a wide range of extensions, such as adding covariates for the dependence parameter of the bivariate copulas as well as extending to multivariate response data with a single set of state variables or separate sets of state variables. Here the use of vine copulas can be envisioned wherever higher-dimensional than bivariate copulas are needed. Another route of extension would be to model the bivariate copulas completely nonparameteric. In this case the identifiability issues have to be reworked.

\section*{Acknowledgements}
The second author was supported by a Global Challenges for Women in Math Science Entrepreneurial Programme grant for a project entitled ``Bayesian Analysis of State Space Factor Copula Models'' provided by the Technical University of Munich. The third author is supported by  the  German  Research  Foundation  (DFG grant CZ 86/4-1). Computations were performed on a Linux cluster supported by DFG grant INST 95/919-1 FUGG.

\section{Contour plots of bivariate copula densities}\label{app}

\begin{figure}[H]
\centerline{%
\includegraphics[trim={0 6cm 0 0},width=1.0\textwidth]{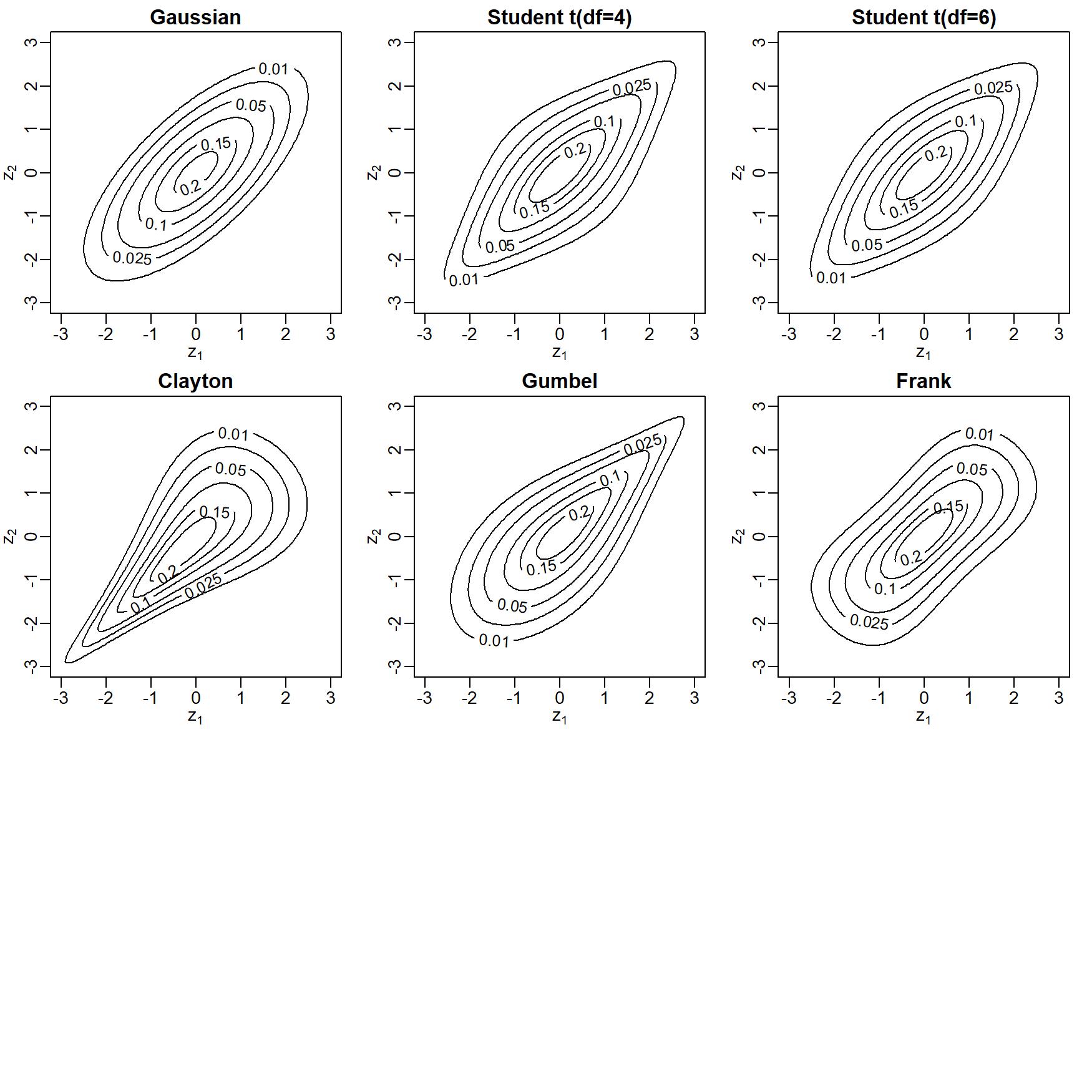}%
}%
\caption{Normalized contour plots of bivariate copula families with Kendall's $\tau = 0.5$.}\label{fig:cplots}
\end{figure}

\section{Hamiltonian Monte Carlo}\label{appendix_hmc}
This section is based on \cite{neal2011mcmc}.
In HMC, our parameters of interest 
are interpreted as a position vector $\boldsymbol q \in \mathbb{R}^d$ at time $s$. Furthermore we assign an associated momentum vector $\boldsymbol p \in \mathbb{R}^d$ at time $s$. The change of the position vector and the momentum vector over time is described through the function $H(\boldsymbol p,\boldsymbol q)$, the \emph{Hamiltonian}, which satisfies the differential equations:
\begin{equation}
\begin{split}
\frac{dq_i}{ds}&=\frac{dH}{dp_i} \\
\frac{dp_i}{ds}&=-\frac{dH}{dq_i}, i=1, \ldots, d.
\end{split}
\label{eq:hdyn}
 \end{equation}
Here we assume that $H(\boldsymbol q, \boldsymbol p) = -\pi(\boldsymbol q|D) + \boldsymbol p^t M^{-1}  \boldsymbol p / 2$,  where $M \in \mathbb{R}^{d \times d}$ is a covariance matrix and $\pi(\boldsymbol q|D)$ is the posterior density for given data $D$. The Leapfrog method is a popular choice to approximate the solution of the differential equations in \eqref{eq:hdyn}, which usually cannot be obtained analytically (\cite{neal2011mcmc}). For our application, the data are the approximately uniform $\hat{u}_1,  \ldots, \hat{u}_{T}$, obtained as in \eqref{eq:uresidual} and the parameter vector $\boldsymbol q$ is given by $\boldsymbol q = (\tau_{lat}, v_1, \ldots, v_T)$. 
Note that for the Bayesian approach the latent variables of the state equation are considered as parameters.
The posterior density is obtained as
\begin{equation*}
\begin{split}
 \pi(\boldsymbol q|D) = \prod_{t=1}^T c_{U,V}(\hat u_t,v_t;\tau_{obs}) \prod_{t=2}^T c_{V_2,V_1}(v_t,v_{t-1};\tau_{lat}),
\end{split}
\end{equation*}
where we assume a uniform prior on the interval (0,1) for $\tau_{lat}$ as specified in Section~\ref{sec:postinf} and  $\tau_{obs}$ is a function of $\tau_{lat}$ as given in \eqref{eq:idrel}.

In order to incorporate the function $H$ into a probabilistic framework, a probability distribution can be defined through the canonical distribution. The corresponding canonical density is given by
\begin{equation}
p(\boldsymbol q,\boldsymbol p) : = \frac{1}{Z} \exp(-H(\boldsymbol p,\boldsymbol q)) = \frac{1}{Z} \pi(\boldsymbol q|D)\exp(-\boldsymbol p^t M^{-1} \boldsymbol p / 2)),
\label{eq:candens}
\end{equation}
where $\boldsymbol q$ and $\boldsymbol p$ are independent and $Z$ is a normalizing constant. Hence, the marginal distribution for $\boldsymbol q$ of $p(\boldsymbol q,\boldsymbol p)$ in \eqref{eq:candens} is the desired posterior distribution. Note that the marginal distribution for $\boldsymbol p$ is a multivariate normal distribution with zero mean and covariance matrix $M$. To sample $\boldsymbol q$ and $\boldsymbol p$ from the canonical distribution specified in \eqref{eq:candens} we proceed as follows. 
\begin{enumerate}
\item Sample $\boldsymbol p$ from the normal distribution with zero mean vector and covariance matrix $M$.
\item Metropolis update: start with the current state $(\boldsymbol q,\boldsymbol p)$ and use the Leapfrog method to simulate L steps of Hamiltonian dynamics  with step size $\epsilon$. We obtain a new state $(\boldsymbol q',\boldsymbol p')$ and accept this proposal with Metropolis acceptance probability
\begin{equation*}
 \min\left(1,\frac{\pi(\boldsymbol q'|D)    \exp(\boldsymbol p^t M^{-1}\boldsymbol p/2)}{\pi(\boldsymbol q|D)    \exp(\boldsymbol p'^t M^{-1}\boldsymbol p'/2)}\right).
\end{equation*}
\end{enumerate}

In conventional HMC, $\epsilon$, $L$ and $M$ need to be specified by the user. The No-U-Turn sampler sets these tuning parameters adaptively during sampling.

\bibliographystyle{spbasic}    
\bibliography{References}{}

\end{document}